\documentclass[12pt]{article}

\usepackage{times}
\usepackage[numbers,sort&compress]{natbib}
\usepackage{graphicx}

\usepackage{ulem}

\usepackage{multirow}

\usepackage[fleqn]{amsmath}
\usepackage{amssymb}

\usepackage[online, flushleft, referable]{threeparttablex}

\usepackage{color}

\usepackage{lineno}

\makeatletter
\renewcommand\@cite[1]{$^{\scriptsize{#1}}$}
\makeatother

\usepackage[tracking=true]{microtype}

\newcommand\normalstyle{\SetTracking{encoding=*}{0}\lsstyle}

\newenvironment{sciabstract}{%
	\begin{quote} \bf}
	{\end{quote}}

\title{Mechanistic Origin of High Strength in Refractory BCC High Entropy Alloys up to 1900K}

\author
{Francesco Maresca$^{\ast}$, William A. Curtin\\
	\\
	\normalsize{Laboratory for Multiscale Mechanics Modeling, Institute of Mechanical Engineering,}\\
	\normalsize{\'{E}cole Polytechnique F\'{e}d\'{e}rale de Lausanne, Lausanne CH-1015, Switzerland}\\
	\\
	\normalsize{$^\ast$E-mail: francesco.maresca@epfl.ch}
}

\date{}

\begin{document} 
	
	
	
	
	\maketitle 
	
	
	
	\begin{sciabstract}
		
The body centered cubic (BCC) high entropy alloys MoNbTaW and MoNbTaVW show exceptional strength retention up to 1900K.  The mechanistic origin of the retained strength is unknown yet is crucial for finding the best alloys across the immense space of BCC HEA compositions. Experiments on Nb-Mo, Fe-Si and Ti-Zr-Nb alloys report decreased mobility of edge dislocations, motivating a theory of strengthening of edge dislocations in BCC alloys. Unlike pure BCC metals and dilute alloys that are controlled by screw dislocation motion at low temperatures, the strength of BCC HEAs can be controlled by edge dislocations, and especially at high temperatures, due to the barriers created for edge glide through the random field of solutes.  A parameter-free theory for edge motion in BCC alloys qualitatively and quantitatively captures the strength versus temperature for the MoNbTaW and MoNbTaVW alloys.  A reduced analytic version of the theory then enables screening over $>$600,000 compositions in the Mo-Nb-Ta-V-W family, identifying promising new compositions with high retained strength and/or reduced mass density. Overall, the theory reveals an unexpected mechanism responsible for high temperature strength in BCC alloys and paves the way for theory-guided design of stronger high entropy alloys.
		
	\end{sciabstract}

\section{Introduction}

\begin{figure*}[h!!]
	\centering
	\vspace*{-0.3cm}
	\hspace*{-0.2cm}
	\includegraphics[width=11.5 cm]{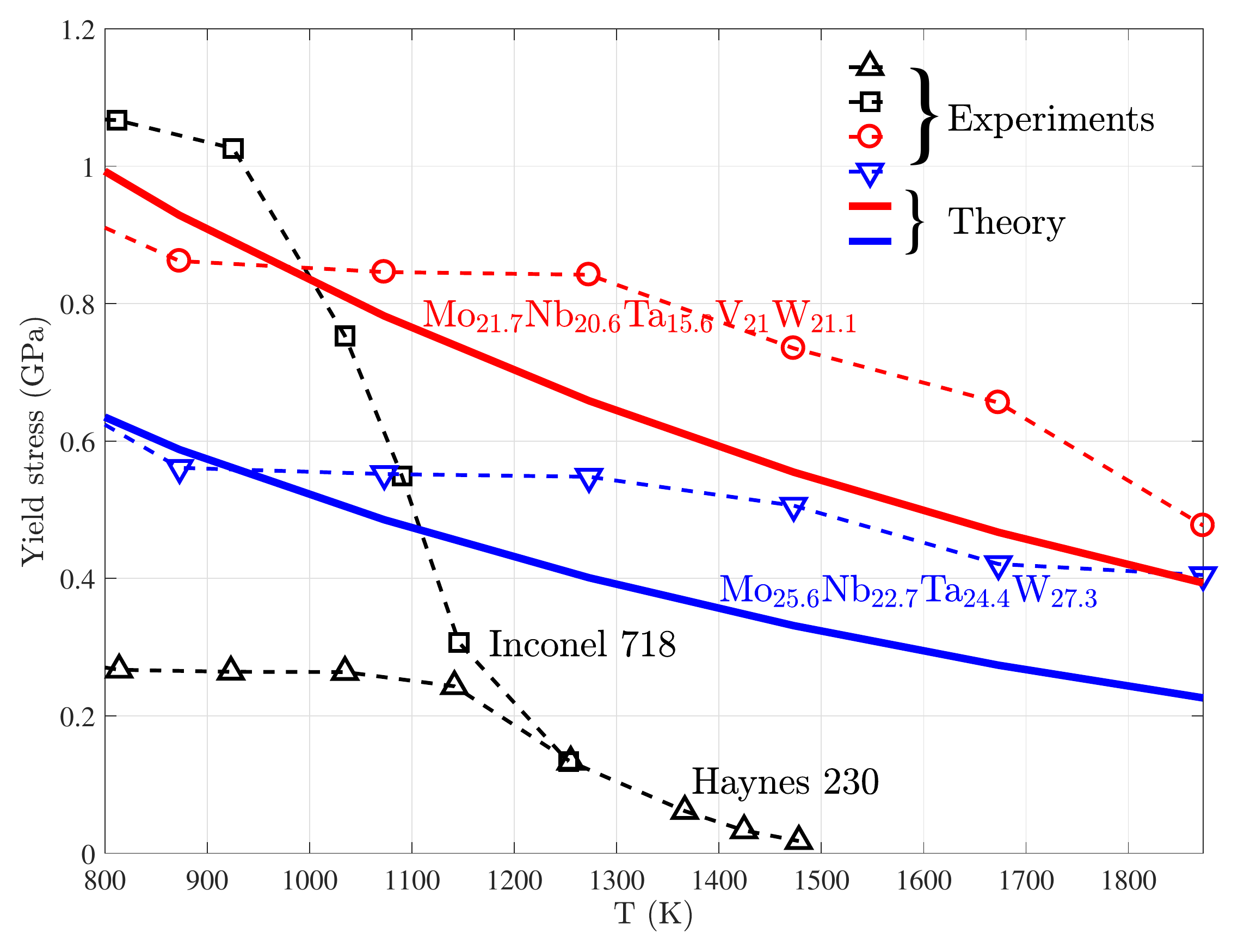}
	\vspace*{-0.7cm}
	\caption{{\bf Strength {\it vs} temperature of BCC HEAs.} Yield strength of BCC high entropy alloys and Ni-based superalloys at high temperatures 800-1900K, showing exceptional strength retention in these HEAs.  Open symbols indicate experiments \cite{Senkov2011} and solid lines indicate predictions.}
	\label{fig1}
\end{figure*}

\vspace*{-0.2cm}
Among the emerging class of ``high entropy alloys'' \cite{Gludovatz2014,Wu2014a,Senkov2010,Senkov2011,Yao2016a,Yao2016b,Li2016}, the body-centered-cubic (BCC) HEAs of nominal compositions MoNbTaW and MoNbTaVW have recently been shown to possess exceptional strengths at temperatures up to 1900K \cite{Senkov2010,Senkov2011}, far above the limits of $\sim$1100K for existing superalloys (Figure \ref{fig1}).

\vspace*{-0.3cm}
These BCC-HEAs consist of the refractory elements Mo, Nb, Ta, V, and/or W at near equal concentrations with the different atom types occupying the crystalline BCC lattice sites at random \cite{Miracle2017}.  The underlying physical origins of this enabling behavior in these high-complexity alloys are unknown.  It has been recently highlighted \cite{Senkov2018} that ``the relationship between composition, microstructure, and high-temperature mechanical properties needs to be established, which can be quite different from the relationships acquired at RT''.  Here, we establish an important component of this relationship, which enables the possibility of discovering new compositions with even better high-temperature performance. 

\begin{figure*}[h!!]
	\centering
	\vspace*{-0.4cm}
	\hspace*{-1cm}
	\includegraphics[width=17.5 cm]{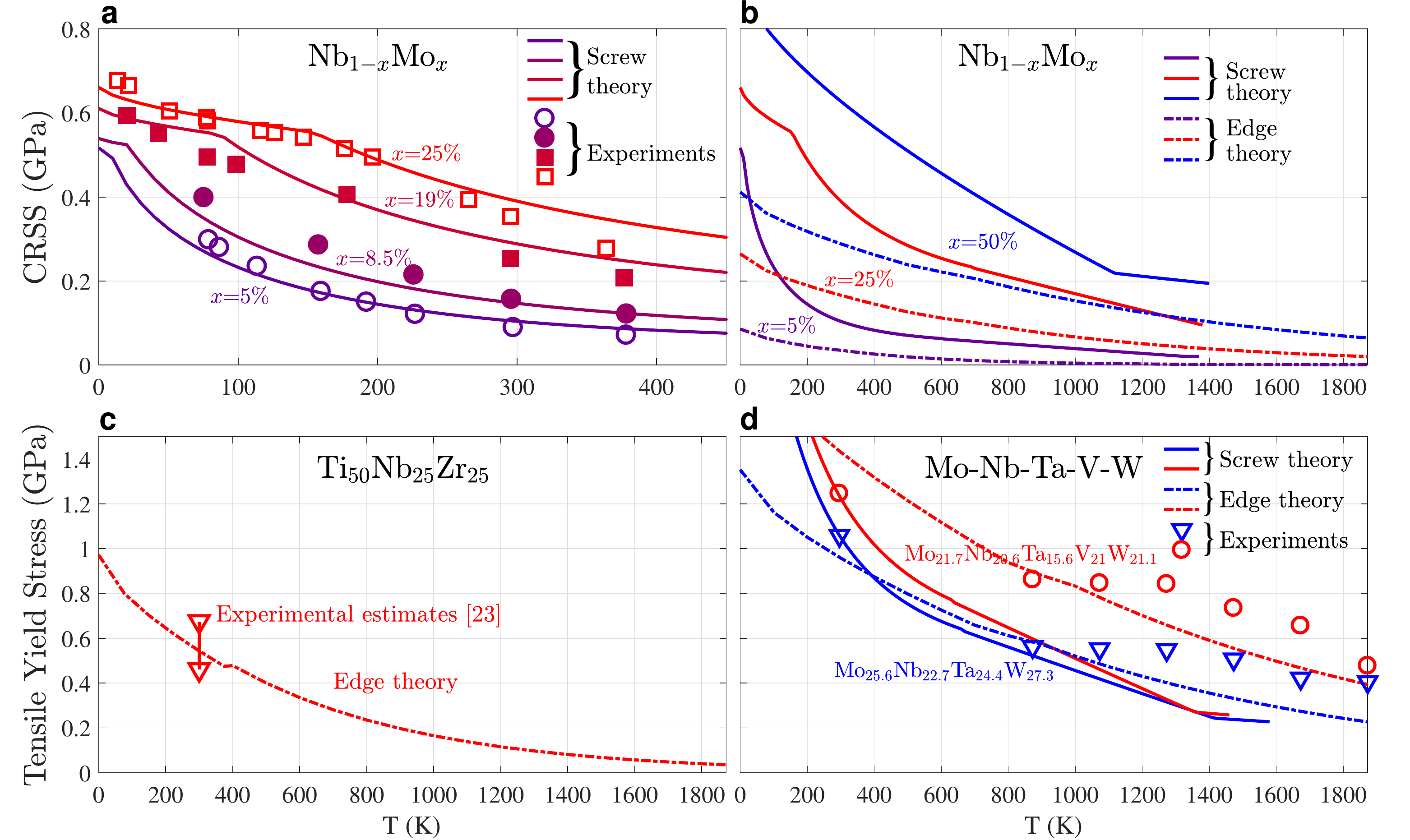}
	\vspace*{-0.5cm}
	\caption{{\bf Screw {\it vs} edge dislocation strengthening in BCC alloys.} \textbf{a} Experimental measurements and predictions of the critical resolved shear strengths (CRSS) {\it vs} temperature in Nb$_{1-x}$Mo$_x$ up to $x=25\%$ (experiments from Ref.~\cite{Statham1972}). \textbf{b} Edge and screw theory predictions for strength {\it vs} temperature in Nb$_{1-x}$Mo$_x$ up to x=50\%. \textbf{c} Edge theory prediction of strength for Ti$_{50}$Nb$_{25}$Zr$_{25}$ versus temperature, with experimental estimates at room temperature also shown \cite{Dirras2017}. \textbf{d} Edge and screw theory predictions and experimental strengths {\it vs} temperature for the NbMoTaW and NbMoTaVW alloys \cite{Senkov2011}; screw theory is terminated at the estimated $T_m/2$ ($T_m$ is the melting temperature) where other mechanisms enter to defeat screw strengthening.}
	\label{fig2New}
\end{figure*}

The yield strength of BCC pure metals \cite{Rodney2014} is well-understood in terms of the motion of screw dislocations via thermally-activated double-kink nucleation.  While the strength is very high (1-2 GPa) at T=0K due to the nucleation barrier, it decreases quickly \cite{Cordero2016} to 100-200 MPa at T$\simeq$300K.  In low-to-moderate concentration binary alloys, the strength increases significantly at low T because, although double-kink nucleation is easier, kink glide becomes strongly inhibited \cite{Suzuki1979,Trinkle2005}.  Glide on different available glide planes also leads to strengthening via jog/dipole formation.  These features are contained within a new theory for screw motion in BCC alloys of arbitrary complexity \cite{Maresca2018b}, and also the classical screw model of Suzuki~\cite{Suzuki1979}, capturing experimental trends.  Figure \ref{fig2New}a shows the predicted critical resolved shear strength (CRSS) for screw motion in Nb$_{1-x}$Mo$_{x}$ versus temperature up to $x=25\%$ along with the data of Statham et al.~\cite{Statham1972} using our new screw theory; the agreement is very good.

TEM observations in both Nb$_{1-x}$Mo$_{x}$ and the classic Fe$_{1-x}$Si$_{x}$, $x\leq 9\%$ alloy indeed reveal that the strength is controlled by screw motion.  However, both Statham et al.~\cite{Statham1972} and Caillard~\cite{Caillard2013} report that edge dislocations have decreasing mobility (higher strength) with increasing solute content.  Quantitatively, Caillard~\cite{Caillard2013} even estimates a CRSS of 125 MPa for the edge dislocation in Fe-9\%Si as compared to 200 MPa for the screw dislocation.  Figure \ref{fig2New}b shows CRSS predictions of our new edge theory, introduced in this paper, to Nb$_{1-x}$Mo$_{x}$ up to $x=50\%$, showing that the edge strength is indeed increasingly competitive with the screw strength with increasing temperature and concentration.  Theory is thus fully consistent with the conclusions of Statham et al.~\cite{Statham1972} in Nb$_{1-x}$Mo$_{x}$.

Recent studies on HEAs in the BCC Ti-Zr-Hf-Nb-Ta family also provide evidence of the increasing importance of edge dislocations.  X-ray line analysis at low plastic strains in TiZrHfNbTa~\cite{Dirras2015} indicates dominance of edge dislocations at the start of plastic flow.  TEM studies~\cite{Couzinie2015} show screw dislocation dominance at larger plastic strains, but new observations at elevated temperatures ($\sim$773K)~\cite{Couzinie2018} show dislocations with considerable curvature and a ``viscous" motion, indicating a loss of strong screw dominance (although the jogs on the screws remain prominent).  In new work on Ti$_{\mathrm{50}}$Zr$_{\mathrm{25}}$Nb$_{\mathrm{25}}$ at room temperature \cite{Mompiou2018}, Mompiou et al. show that edges are also sluggish, becoming comparable in strength to screw dislocations.  Specifically, they measure a velocity of 28.5 nm/s for edge dislocations and 4.5 nm/s for screw dislocations, a difference that corresponds to only a small difference in stress levels needed to drive the dislocations.  Predictions of our new theory for edge strength in Ti$_{\mathrm{50}}$Zr$_{\mathrm{25}}$Nb$_{\mathrm{25}}$ are shown in Figure \ref{fig2New}c, and are within the range of experiments indicating that edge strength is comparable to the screw strength.

Other evidence for the role of the edge dislocation in the strengths of BCC HEAs also exists.  Yield strengths have been correlated with solute misfit volumes \cite{Chen2018,Yao2017}; this is a hallmark of edge-dominated strengthening.  Furthermore, first-principles and interatomic potential computations of solute/screw interaction energy in the refractory metals show no correlation with solute misfit volumes.  To explain other experiments in BCC alloys, simplified versions of the Suzuki model \cite{Argon2007,Trinkle2005} invoke an athermal stress operative at elevated temperatures and attributed to a vague ``solute pinning''; this is precisely what the transition to edge dislocation dominance achieves.  Thus, while never previously considered as relevant in BCC alloys, there is unambiguous experimental support in both old and new literature for the emergence of edge dislocation motion as important in high concentration/complex BCC alloys especially at higher temperatures.

We now apply the same new screw and new edge theories to the HEA alloys MoNbTaW and MoNbTaWV (Figure~\ref{fig2New}d).   The screw theory material parameters are fit to match the uniaxial tensile experiments at T=300K.  Then, although both (fitted) screw and (parameter-free) edge theories agree with experiment at T=300K, the screw strength is far lower than experiments at higher T.  Edge dominance emerges at moderate temperatures and accurately predicts the high-temperature behavior in these alloys (also see Figure \ref{fig1}).  Furthermore, the screw theory predicts almost no difference in performance between MoNbTaW and MoNbTaWV while experiments and the edge theory both show that the 5-component alloy containing V is notably stronger than the 4-component alloy.  The latter result is surprising since V has the lowest melting point among all the constituent elements, which would normally suggest that V-containing alloys have lower strengths at high T.  The high strength of the edge dislocation, and its dominance over the screw dislocation at higher temperatures, is unexpected.  But these results show that the edge dislocation is essential for understanding the high retained strength in the MoNbTaW and MoNbTaWV HEA alloys.

Having introduced the role of edge dislocations in strengthening of BCC alloys, we proceed with detailed analysis in the remainder of this paper. In Section 2 we present the theory of edge dislocation strengthening in BCC random alloys. In Section 3, the theory is validated against Molecular Statics simulations at T=0K. In Section 4, we compare predictions of the edge model to experiments on Nb-Mo-Ta-W-V high entropy alloys in more detail. In Section 5, we use the theory to search over $>$600,000 compositions in the Mo-Nb-Ta-V-W family and we identify new alloys predicted to have even higher strength or strength/weight ratios.  Many other compositions with comparable strengths are predicted, and can satisfy additional performance requirements.  Overall, these insights and theory open a new direction for theory-guided design of advanced high-temperature materials based on the high-entropy concept.

\vspace*{-0.5cm}
\section{Theory of edge dislocation motion in BCC alloys}

\vspace*{-0.2cm}
The theory of strengthening for edge dislocations in BCC high entropy alloys follows conceptually the analyses previously presented for dilute and HEA FCC alloys \cite{Leyson2012,Varvenne2016}. The random distribution of solutes in the lattice leads to local fluctuations in the solute concentrations.  The dislocation is attracted to fluctuations that lower the system energy and is repelled by fluctuations that increase the system energy.   A long edge dislocation line, which unlike the screw dislocation is very flexible, therefore adopts a wavy configuration as it finds energetically-favorable regions of solutes.  The waviness is constrained by the energy cost of increasing the dislocation line length and curvature (i.e. constrained by line tension).  A characteristic waviness thus emerges, which is characterized by an amplitude $w_c$ and a lateral length $\zeta_c$, with the wavelength some multiple of $\zeta_c$, which is the scale at which the total system energy is minimized.  In the minimum energy state, dislocation segments of length $\zeta_c$ reside in local minimum energy positions (locally favorable solute fluctuations).  $w_c$ is the distance between consecutive local minima and maxima of the energy fluctuation.  Dislocation motion occurs by thermal activation of the $\zeta_c$ segments residing in the local minima over the adjacent local maxima at distance of $w_c$.  An applied resolved shear stress reduces the barrier for thermal activation, and the zero-temperature flow stress is the stress at which the barrier is zero such that athermal motion can occur.  We now provide the details of the analysis sketched above.  The analysis here is much more detailed than the original analysis of Leyson et al. for FCC alloys, leading to some numerical differences but with the same overall scalings of strength and energy barrier with underlying material properties.

We consider a general $N$-component alloy with concentration $c_n$ of the $n^{th}$ element ($\sum_{n=1}^N c_n =1$).  We then envision the edge dislocation as existing in a homogeneous ``average" alloy that is the effective ``matrix" for the true random alloy.  Every individual atom is then considered as a solute in the average matrix - the alloy is thus effectively at 100\% solute concentration.  The solutes ($n=1,...N$) have, for instance, misfit volumes {$\Delta V_n$} in the average alloy matrix.  These solutes interact with the dislocation in the average matrix.  The interaction energy between the dislocation, centered at the origin and aligned along $z$, and a solute of type $n$ at position $x_i, y_j$ is denoted as $U_n (x_i, y_j)$ (e.g. Fig. \ref{fig2}b for Nb in $\overline{\mathrm{NbTaV}}$).  For dislocation glide in random alloy, the key quantity is the energy change of a straight dislocation segment of length $\zeta$ upon glide by a distance $w$.  Due to the specific solute arrangements, the energy change of the solute/dislocation-segment system differs from the same dislocation segment in the homogeneous alloy by $\Delta U_{\mathrm{tot}}(\zeta,w)$.  This energy change is a random variable \cite{Varvenne2016} with standard deviation given by
\vspace*{-0.5cm}
\begin{equation}
\sigma_{\Delta U_{\mathrm{tot}}} = \left[ \langle \Delta U_{\mathrm{tot}}^2(\zeta,w) \rangle - \langle \Delta U_{\mathrm{tot}}(\zeta,w)\rangle^2 \right]^{\frac{1}{2}} \ .
\end{equation}
Carrying out the averages (see Ref.~\cite{Varvenne2016}) in the random alloy, and  noting that for the BCC edge there are $\zeta/2\sqrt{2}b$ nominally identical sites along $z$ having the same $(x,y)$ position relative to the dislocation line, enables us to write the standard deviation as
\vspace*{-0.2cm}
\begin{equation}
\sigma_{\Delta U_{\mathrm{tot}}} = \Delta E_p (\zeta,w) = \left( \frac{\zeta}{2\sqrt{2} b} \right)^{\frac{1}{2}} \Delta \tilde{E}_p (w)
\label{NEq2}
\end{equation}
\vspace*{-0.5cm}
where
\begin{equation} 
\Delta \tilde{E}_p (w) = \Bigg[ \sum_{i,j,n}  c_n\Big( U_n(x_i-w,y_j) - U_n(x_i,y_j) \Big)^2  \Bigg]^{\frac{1}{2}}\ 
\label{eq:energyscale}
\end{equation}
is the key potential energy contribution per unit Burgers vector of dislocation line length.

\begin{figure*}[h!!]
	\centering
	\includegraphics[width=10.7 cm]{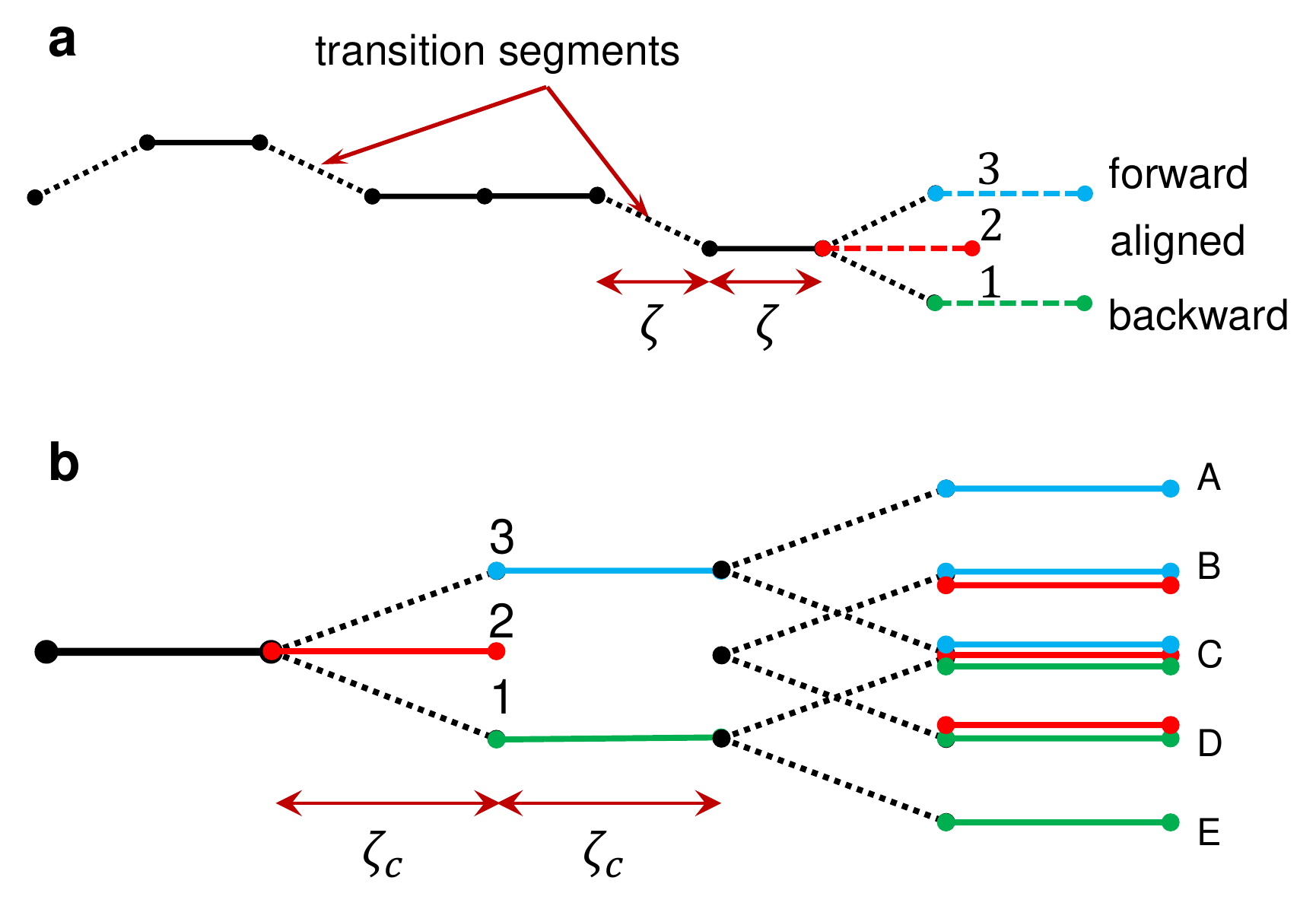}
	\vspace*{-0.4cm}
	\caption{{\bf Three-choice model.} \textbf{a} Sketch of the dislocation line, composed of length $\zeta$ segments connected by transition segments, with three possible positions of each straight segment relative to the previous segment (in-line, ahead, behind). \textbf{b} All possible configurations of two consecutive segments ((1-2-3) and (A-B-C-D-E)) within the three-choice model.}
	\label{fig5}
\end{figure*}

The low-energy structure for the edge dislocation with waviness $w$ and $\zeta$ is then determined as follows.  We start at an arbitrary segment and consider the possible positions and energies for the next segment of length $\zeta$ relative to the previous segment (see Figure~\ref{fig5}a).  The next segment has three possible positions along the glide plane: in-line with the previous segment, and ahead/behind by $w$.  The potential energy of each of these possible positions is randomly chosen from the potential energy distribution having the standard deviation $\Delta E_p(w)$.  If the next segment is ahead or behind, there is a transition segment of lateral length $\zeta$ that makes no contribution to the potential energy but enables the transition in glide position and maintains a smooth wavy dislocation line with a characteristic scale $\zeta$.  A transition segment increases the elastic energy by $E_L = \Gamma \frac{w^2}{2\zeta}$ where $\Gamma$ is the dislocation line tension and $w << \zeta$ is assumed.  The choice of segment (in-line, ahead, behind) is then the segment that lowers the total energy (potential plus elastic).  Note that some fraction of segments remain in-line, and so there is some ratio $\kappa < 1$ of  transition segments to straight segments.  Within the above construction, the total energy of a long dislocation line is then the sum of the potential energies of each straight segment of length $\zeta$ plus the elastic energies due to the transition segments.  The total energy can then be written as
\vspace*{-0.2cm}
\begin{equation}
\Delta E_{\rm tot}(\zeta,w) / L = \left[\kappa\Gamma \frac{w^2}{2\zeta} - \beta \left(\frac{\zeta}{2\sqrt{2}}\right)^{\frac{1}{2}}\Delta \tilde{E}_p(w) \right] \frac{1}{(1+\kappa)\zeta},
\label{eq:E_p_E_line2}
\end{equation}
where $\beta \Delta \tilde{E}_p(w)$ is the average energy reduction due to the straight segments finding favorable solute environments. Both $\kappa$ and $\beta$ must be determined as discussed below.

\vspace*{-0.3cm}
The minimum energy dislocation structure is found by minimizing Eq.~(\ref{eq:E_p_E_line2}) with respect to $\zeta$ and $w$.  Minimization with respect to $\zeta$ is analytic and yields
\vspace*{-0.2cm}
\begin{align}
\zeta_c(w) = 4^{\frac{1}{3}} \left( \frac{\kappa^2 2\sqrt{2}}{\beta^2} \right)^{\frac{1}{3}} \left(\frac{\Gamma^2w^4b}{\Delta \tilde{E}_p^2(w)}\right)^{\frac{1}{3}}.
\label{NEq3}
\end{align}
Substitution of $\zeta_c$ back into the total energy reveals that the typical potential energy gain due to solutes $E_P = \beta \Delta E_p$ is always equal to four times the elastic energy cost $E_L$,
\vspace*{-0.2cm}
\begin{equation}
E_P = 4 E_L \ .
\label{NEq3b}
\end{equation}
\vspace*{-0.8cm}

This relationship is needed below to compute $\kappa$ and $\beta$.
The total energy per unit length can further be written as
\vspace*{-0.2cm}
\begin{equation}
\Delta E_{\rm tot}/L = - \frac{3}{2^{\frac{7}{3}}} \left( \frac{\beta^4}{(2\sqrt{2})^2 \kappa} \right)^{\frac{1}{3}} \left( \frac{\Delta \tilde{E}_p^4(w)}{b^2 w \Gamma} \right)^{\frac{1}{3}} \ .
\label{NEq4}
\end{equation}
\vspace*{-0.5cm}

Minimization with respect to $w$ leads to the characteristic waviness amplitude $w_c$.  This must be done numerically, but reduces to the solution only of $d \Delta \tilde{E}_p(w) / dw = \Delta \tilde{E}_p(w) / 2w$.  $w_c$ therefore does not depend on $\beta$ and $\kappa$. 

The values of $\kappa$ and $\beta$ are calculated numerically following the discrete ``three-choice" model introduced above (Figure~\ref{fig5}).  In selecting the next segment, now of known  length $\zeta_c$, the potential energies of all three possible choices at glide distances $(w_c, 0, -w_c)$ are selected randomly from the distribution ($\beta_i \Delta E_p(w_c)$ where $\beta_i$ $i=1,2,3$ are random numbers chosen from a Gaussian distribution with zero mean and unit standard deviation).  Including the line energy $E_L$ for the ahead/behind segments $i=1,3$, the minimum total energy segment is chosen.  There is thus a bias toward remaining in-line ($i=2$).  This process is then repeated many times (equivalent to moving along the line $\zeta_c$ segment by $\zeta_c$ segment).  Averaging over all the choices leads to the average fraction $1-\kappa$ of in-line segments and the average potential energy $\beta \Delta E_p$ of the selected segments.  However, the line energy cost $E_L$ and potential energy $E_P$ depend on $\kappa$ and $\beta$ but also must satisfy 
$E_P = 4 E_L$, which is independent of $\kappa$ and $\beta$.  A self-consistent procedure is thus used.  We start with the values obtained by assuming no bias effect, $\kappa=2/3$ and $\beta=0.85$, execute the stochastic simulations, obtain a new $\kappa$ value, and then determine a new $\beta$ value by enforcing $E_P = 4 E_L$.  This process is iterated until the values of $\kappa$ and $\beta$ are converged.  Using 500,000 segments, the converged values are $\kappa = 0.56$ and $\beta = 0.83$.  The value of $\kappa$ means that nearly 1/2 of the segments are in-line with another segment, and that the average repeating unit consists of approximately two in-line segments plus one transition segment, with a precise length of 2.8$\zeta_c$.  The waviness of the dislocation is thus characterized by a wavelength of 5.6$\zeta_c$ and amplitude $w_c$.

Having determined the minimum energy configuration of the wavy dislocation, we now determine the typical energy barrier for the glide of the independent segments of length $\zeta_c$.  For a given segment, the stochastic simulation provides the potential energy change upon advance of the segment by glide of distance $w_c$.  This glide can be accompanied by annihilation or creation of transition segments, depending on the configurations of the surrounding segments.  All possible configurations within the three-choice model are shown in Fig. \ref{fig5}b, and the advance of the central segment (1, 2 or 3) is considered. An advance that involves the creation of new transition segments has a high energy barrier, and does not occur.  The entire line thus moves forward locally by those processes that do not create new transition segments.  However, there is only one low energy barrier case (configuration 1-C with segment 1 moving forward by $w_c$) that annihilates two transition segments upon glide.  This case occurs on average, however, with a low probability $\kappa^2/4 = 7.8\%$.  There are no cases that annihilate a single transition segment.  Thus, the dominant processes controlling motion involve advancement with no line energy gain/cost, corresponding to advance of section 2 in case 2-B and advance of section 1 in case 1-D; these constitute a total probability $(1-\kappa)\kappa = 25\%$ of possible cases.  The energy barrier for motion in these cases is then only the potential energy difference between the lower-energy initial state and the higher-energy transition state.  Our three-choice stochastic model gives this difference as $\approx \sqrt{2}\Delta E_p$, consistent with previous direct stochastic simulations of a segment moving through a random field of solutes \cite{Leyson2012}.  Note that new transition segments are formed as segments cross over the transition state and fall into the next low-energy configuration.  Once the entire long dislocation line has advanced by $\approx 2 w_c$, it is in a new statistical low-energy configuration.  Based on the above analysis, the typical energy barrier for advancement of the $\zeta_c$ segments is 
\begin{equation}
\label{eq:DE_b}
\Delta E_b =  \sqrt{2} \left( \frac{\zeta_c}{2\sqrt{2} b} \right)^{\frac{1}{2}} \Delta \tilde{E}_p = 1.11\left( \frac{w_c^2 \Gamma \Delta \tilde{E}_p^2(w_c)}{b}\right)^{\frac{1}{3}} \ .
\end{equation}
The final result above has exactly the same form as derived by Leyson et al., but with a slightly different numerical prefactor (1.11 vs. 1.22) that arises from differences in analysis (Leyson et al. essentially have $\kappa = 1$, $\beta=1$) and BCC vs. FCC atomic spacing along the edge line direction ($2\sqrt{2}b$ and $\sqrt{3}b$, respectively).

To glide, the dislocation must overcome the barrier $\Delta E_b$ by thermal activation but assisted by the work $ -\tau b \zeta_c x$ done by an applied resolved stress $\tau$ on the length $\zeta_c$ segment as it glides a distance $x$ relative to the minimum energy position.  For a sinusoidal energy landscape, the stress-dependent energy barrier is$^{26}$
\begin{equation}
\label{eq:DEb_tau}
\Delta E(\tau) = \Delta E_b \left( 1 - \frac{\tau}{\tau_{y0}} \right)^{\frac{3}{2}} 
\end{equation}
where $\tau_{y0}$ is the zero-temperature flow stress, given as
\vspace{-0.2cm}
\begin{equation}
\tau_{y0}  = \frac{\pi}{2}\frac{\Delta E_b}{b\zeta_c(w_c)w_c}  =  1.01  \left(\frac{\Delta \tilde{E}_p^4(w_c)}{\Gamma b^5 w_c^5} \right)^{\frac{1}{3}}.
\label{eq:tau_y0}
\end{equation}
The above result is exactly the same as found for FCC by Leyson et al., including the numerical prefactor.  This is fortunate but fortuitous, stemming from cancellation among the different numerical factors in the theory. 

At stresses $\tau < \tau_{y0}$, and for quasi-static loading, the plastic strain-rate $\dot{\varepsilon}$ is related to the energy barrier through a thermally-activated Arrhenius model~\cite{Argon2007,Kocks1975}  $\dot{\varepsilon}=\dot{\varepsilon}_0\exp \left(-\Delta E(\tau)/kT\right)$. Combining this with Eq.~(\ref{eq:DEb_tau}), leads to the finite-temperature, finite strain-rate flow stress $\tau_y (T,\dot{\varepsilon})$ as
\vspace*{-0.3cm}
\begin{equation}
\tau_y(T, \dot{\varepsilon}) =  \tau_{y0}\left[1-\left(\frac{kT}{\Delta E_b}\ln \frac{\dot{\varepsilon}_0}{\dot{\varepsilon}}\right)^{\frac{2}{3}}\right]\ ,
\label{eq:tau_y_T_epsdot2}
\end{equation}
which holds for low temperatures and high stress ($\tau_y/\tau_{y0} > 0.5$). Here, $\dot{\varepsilon}_0$ is a reference strain-rate estimated as $\dot{\varepsilon}_0=10^{4}$s$^{-1}$ (\cite{Varvenne2016}).  For higher temperatures/lower stress ($\tau_y/\tau_{y0} < 0.5$), the dislocation can explore higher wavelengths and hence the following relation holds~\cite{Leyson2009}
\begin{equation}
\tau_y(T,\dot{\varepsilon})=\tau_{y0} \exp \left(-\frac{1}{0.55} \frac{kT}{\Delta E_b} \ln \frac{\dot{\varepsilon}_0}{\dot{\varepsilon}} \right)\ , \tau_y / \tau_{y0} < 0.5\ .
\label{eq:finiteT_tay_y2}
\end{equation}
The theory thus also predicts an activation volume $V = \frac{3\Delta E_b}{3\tau_{y0}}  (\frac{kT}{\Delta E_b} \ln \frac{\dot{\varepsilon}_0}{\dot{\varepsilon}})^{1/3} \sim w_c \zeta_c b$ which reflects directly the underlying material length scales ($\zeta_c, w_c$) of the wavy dislocation.

We discuss inputs to the theory and a reduced elasticity model in later sections of the paper.

\section{Theory validation against Molecular Statics simulations}
	
\begin{figure*}[h!!]
	\centering
	\hspace*{0cm}
	\includegraphics[width=16 cm]{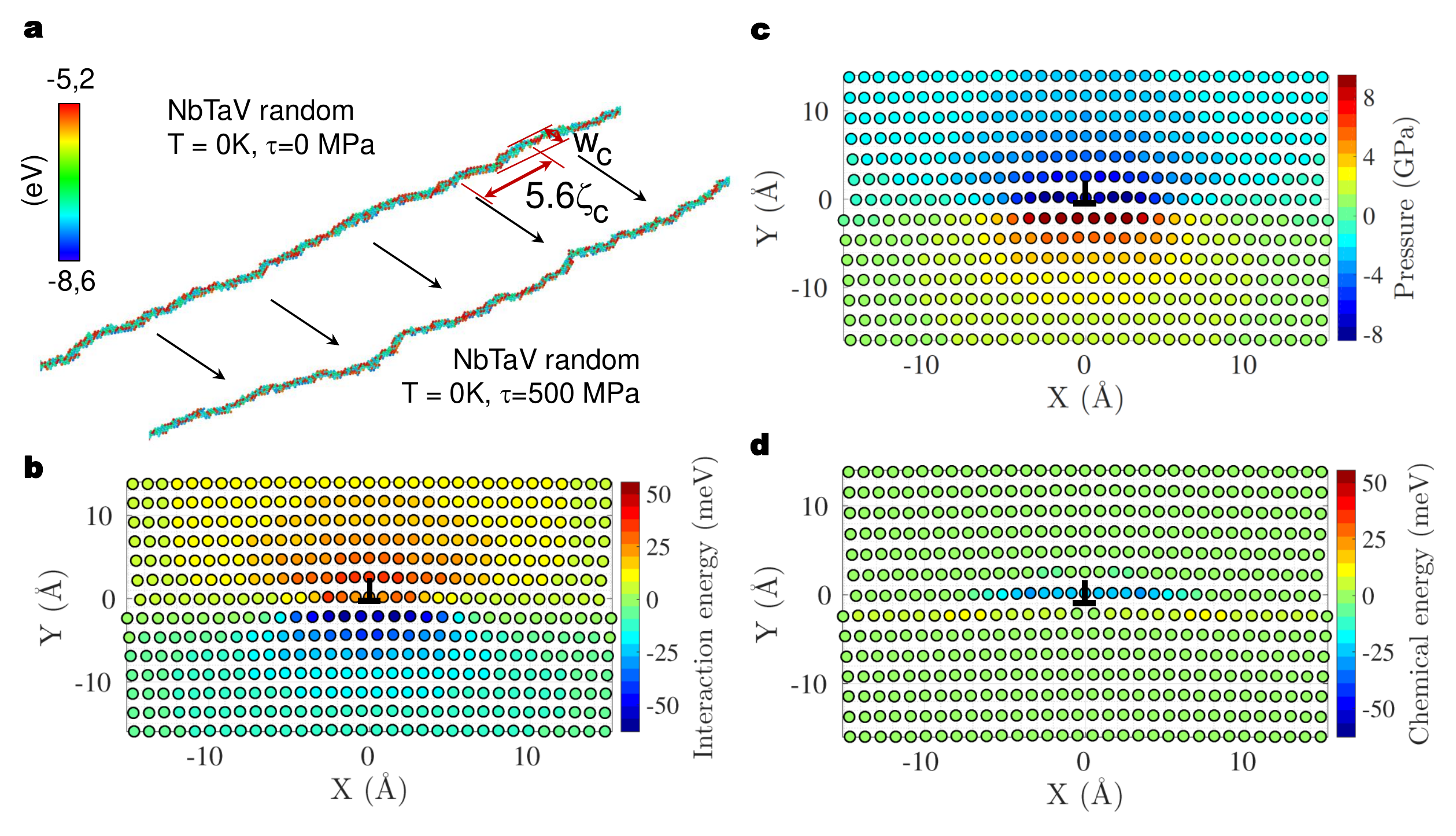}
	\vspace*{-1.3cm}
	\caption{{\bf Edge dislocations in the average and random BCC NbTaV alloy and solute/edge-dislocation interactions.} \textbf{a} (top) Zero temperature and zero stress low-energy wavy configuration of the BCC edge dislocation in the true random NbTaV alloy and (bottom) the dislocation configuration in the true random alloy after some gliding at the critical resolved shear stress $\tau\simeq500$~MPa (corresponding to tensile yield stress $M\tau\simeq 1.5$~GPa). \textbf{b} Solute/edge dislocation interaction energy $U_{\mathrm{Nb}}(x_i,y_j)$ for a Nb solute in the $\overline{\mathrm{NbTaV}}$ material. \textbf{c} Pressure field of the edge dislocation in the $\overline{\mathrm{NbTaV}}$ material.  \textbf{d} Chemical contribution to the solute/edge dislocation interaction energy $U_{\mathrm{Nb}}^{\mathrm{chem,EAM}}$ in the $\overline{\mathrm{NbTaV}}$ alloy (see text). Crystallographic visualizations use OVITO~\cite{Stukowski2010}.}
	\label{fig2}
\end{figure*}

We now validate the edge theory against atomistic simulations of edge motion at T=0K on alloys in the Mo-Nb-Ta-V-W family.  All Molecular Statics simulations have been performed with the LAMMPS package~\cite{Plimpton1995}.   We use EAM-type interatomic potentials~\cite{Zhou2004,Lin2013,Rao2017} for these elements and their alloys.  Since we are studying the edge dislocation, the usual challenges in the application of EAM-type potentials to screw core structures is not relevant.  In comparing theory to simulation, the precision of the potentials with respect to real alloys is also not essential.
Potentials having the EAM form are amenable to a mathematical homogenization method that creates a single ``average atom" potential for the  effective matrix of a given alloy \cite{Varvenne2016b}.  The edge dislocation core structures in the pure elements and in all alloys studied here show no dissociation and spreading of the core over a width of a few Burgers vectors, as expected based on Peierls-Nabarro-type models of dislocation core structures.  As an example, the projected core structure for the average-atom $\overline{\mathrm{NbTaV}}$ is shown in Figure \ref{fig2}.  

To simulate edge dislocation glide, we use the Periodic Array of Dislocations (PAD) configuration~\cite{Bacon2009}.  Samples are oriented with glide direction X$\|[\bar{1}11]$, glide plane normal Y$\|[101]$ and line direction Z$\|[12\bar{1}]$ of dimensions X=56~nm, Y=14~nm and Z=110~nm ($\sim5\cdot10^6$ atoms). Very long dislocation lines ($>> \zeta_c$) are needed to capture the fluctuations in dislocation motion in the random alloy.  Periodic boundary conditions are imposed along X and Z, and zero tractions are imposed on the surfaces normal to the Y direction. A perfect BCC lattice is then populated with the average atoms at the desired alloy composition. An edge dislocation with line direction along Z is introduced by removing two atomic planes with normal along X sitting in the lower half of the simulation cell, and by imposing a linear displacement $u_X = bx/l_x$ for $0<x<l_x$ on all atoms in the lower half of the simulations cell. Atomic positions are then relaxed by using a combination of the FIRE algorithm~\cite{Bitzek2006} and relaxation of the cell dimensions until convergence is achieved (force tolerance $10^{-6}$ eV/atom and stresses $\sigma_{XX}$, $\sigma_{XZ}$ and $\sigma_{ZZ}<1$ MPa). The true random alloy with the dislocation is then created by replacing the average atom at each BCC lattice site with the solute atoms of the alloy chosen at random according to the solute concentrations.  The system is then relaxed to a minimum-energy configuration with the same tolerances on forces and stresses as specified above. As predicted by the theory, the relaxed equilibrium T=0K edge dislocation configuration is spontaneously wavy on the $\{110\}$ planes with waviness amplitude $\sim w_c$, see Table \ref{TableS1} further below. An external stress is then applied by assigning forces to the upper and lower Y boundary atoms over a thickness of few atomic layers. For a desired applied stress $\tau$, the magnitude of the forces \textbf{f} and $-$\textbf{f} applied on the top and bottom surfaces along the X direction is $f = \frac{\tau l_x l_z}{n}$ where $n$ is the number of boundary atoms on the top or bottom layer. The T=0K yield strength is computed by applying stress in 25 MPa increments and relaxing the system with FIRE algorithm and tolerance specified above. The yield stress is the stress at which the dislocation is observed to have moved a distance $w_c/2$ or larger from its original position, along the entire dislocation length.

The theory requires the solute-dislocation interaction energies.  These are computed using the average-alloy dislocation cores in simulation cells with same orientation and boundary conditions as specified above, and dimensions X=14~nm, Y=14~nm and Z=3~nm ($36,000$ atoms).  The creation of the edge dislocation and subsequent cell relaxation procedure is the same as specified above.  Each elemental atom (solute) is then inserted into each possible unique atomic position $(x_i, y_j)$ around the dislocation (within one periodic length along $z$).  The interaction energy $U_n(x_i,y_j)$ for solute $n$ is computed by measuring the energy of the fully-relaxed cell and subtracting the energy of a single solute in an infinite perfect crystal.  We compute the $U_n(x_i,y_j)$ for all solutes at the average alloy compositions listed in Table \ref{Table1}.  Figure \ref{fig2}b shows the interaction energy for Nb solutes in the $\overline{\mathrm{NbTaV}}$ alloy as an example.
	\vspace*{-0.4cm}

From the average-alloy elastic constants $(C_{11},C_{12},C_{44})$, the isotropic alloy shear modulus is computed as $\bar{\mu} =\sqrt{\frac{1}{2}\bar{C}_{44}(\bar{C}_{11}-\bar{C}_{12})}$, the bulk modulus as $\bar{B} = (\bar{C}_{11} + 2\bar{C}_{12})$ and hence $\bar{\nu} = \frac{3\bar{B}-2\mu}{2(3\bar{B}+\mu)}$.  The dislocation line tension can be expressed generally as $\Gamma=\alpha \mu b^2$, and is dominated by elasticity (although there are core energy contributions relevant at very small lengths).  Prior work in FCC materials suggests $\alpha=1/8 - 1/16$ and we use the value $\alpha = 1/12$ here (the dependence of results on $\alpha$ is shown in Appendix \ref{Supp5}).

\begin{figure*}[h!!]
	\centering
	\hspace*{-1cm}
	\includegraphics[width=16 cm]{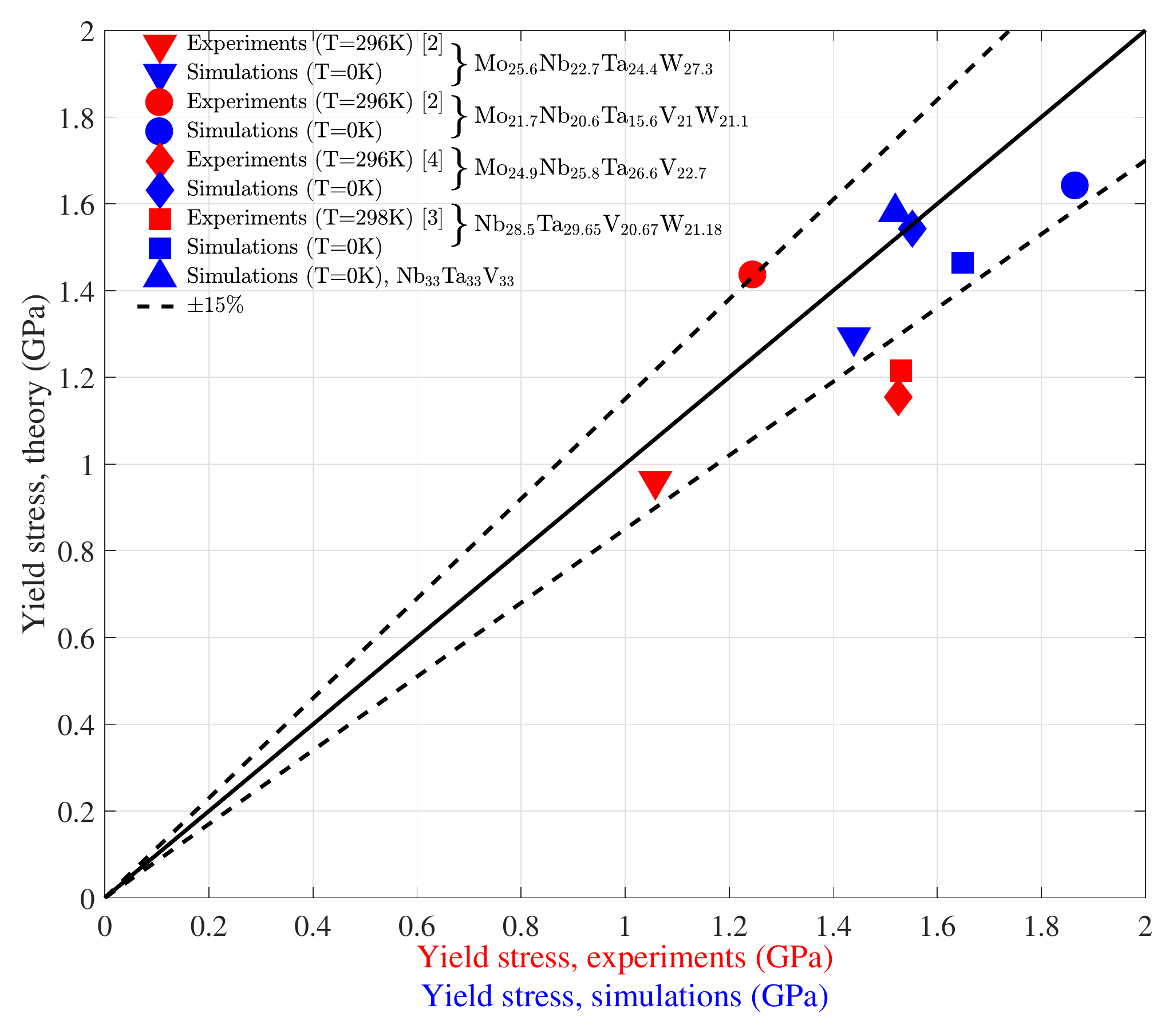}
	\caption{{\bf Theory predictions {\it vs} simulations and experiments}. Predicted yield strength {\it vs} simulations at T=0K (blue symbols), and {\it vs} experiments at room-temperature T$\simeq$300K (red symbols), using the experimental strain rates ($\dot{\varepsilon}=10^{-3}$, Ref.~\cite{Senkov2011}, and $\dot{\varepsilon}=5\cdot10^{-4}$, Refs.~\cite{Yao2016a},\cite{Yao2016b}).}
	\label{fig3}
\end{figure*}

\begin{figure*}[h!!]
	\centering
	\includegraphics[width=15 cm]{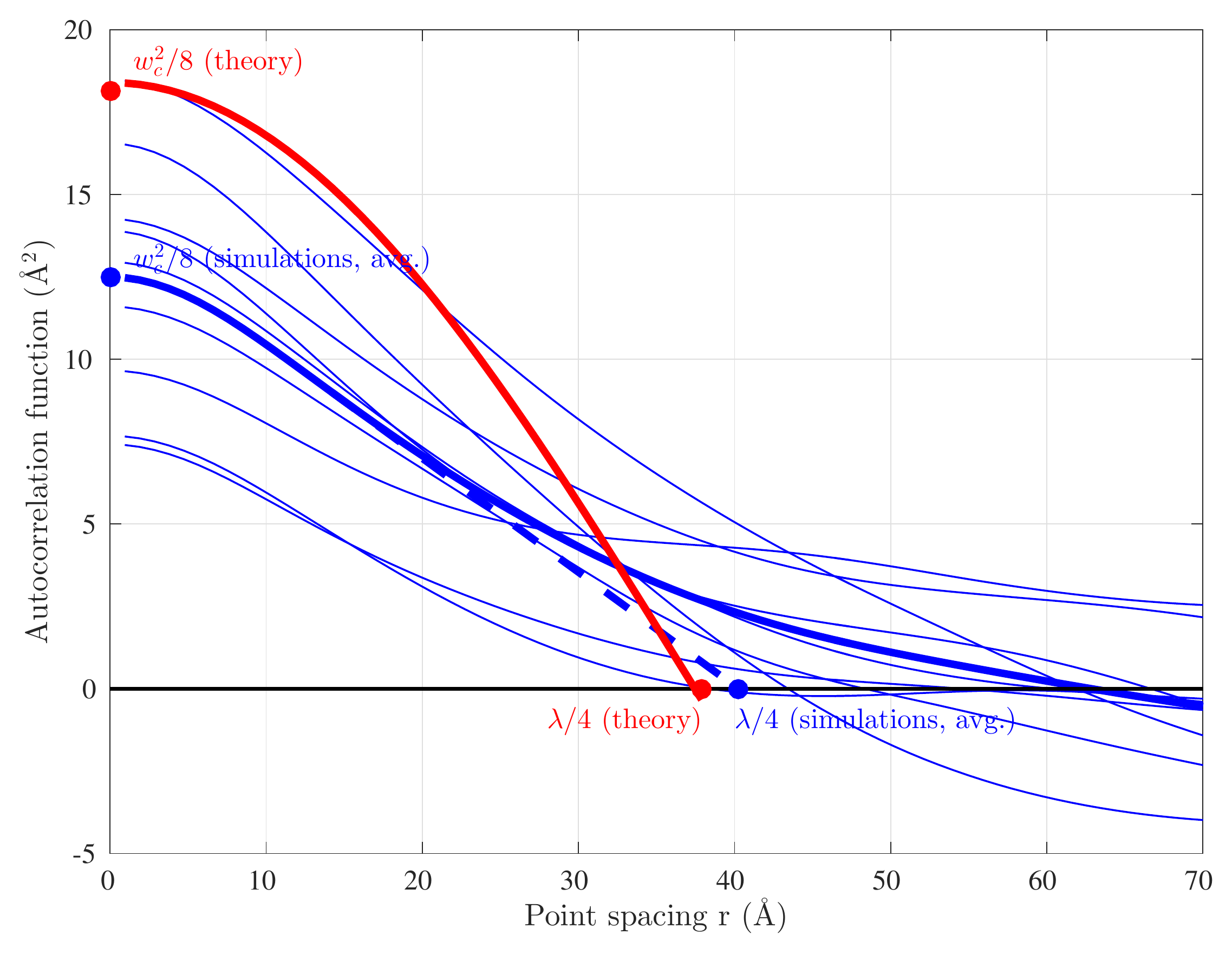}
	\vspace*{-0.5cm}
	\caption{{\bf Characteristic dislocation length scales, theory {\bf vs} simulations.} Height-height correlation function g(r) for 10 relaxed dislocation lines at T=0K for the NbTaV alloy (thin blue lines) and predicted g(r) assuming a sinusoidal configuration (red line).  The thick blue line shows the average g(r) in the simulations and the dashed blue line shows the estimated correlation function corresponding to the sinusoidal model.  The values of the characteristic dislocation length scales $\zeta_c$ and $w_c$ are indicated.  Results are typical of all alloys studied here (see Table \ref{TableS1}).}
	\label{SI11}
\end{figure*}

Figure \ref{fig3} shows the strength predictions at T=0K for a range of alloys versus the simulated strengths.  The agreement, with no adjustable parameters, is very good across all alloys.  We can further compare predicted and estimated values for $w_c$ and $\zeta_c$.  The waviness of the simulated dislocation is characterized by the height-height correlation function $g(r) = \langle x(z)x(z-r)\rangle$ where $x(z)$ is the deviation of the dislocation line from the average dislocation line position $x_{\mathrm{avg}}=\langle x(z) \rangle$.  For a sinusoidal function of amplitude $h$ and wavelength $\lambda$,  $g(0)=h^2/2$ and $g(\lambda/4)=0$.  The wavy dislocation is predicted to have a wavelength $5.6 \zeta_c$ and amplitude $w_c/2$.  For each alloy composition, we measure $g(r)$ for 10 dislocation lines extracted using the DXA algorithm~\cite{Stukowski2010b}.
The measured and computed correlation functions are shown in Figure \ref{SI11} for the NbTaV alloy.  The simulations always show some tail in the correlation function since the dislocation is not globally sinusoidal.  Otherwise, the predicted and measured correlation functions are in good agreement.  The predicted and simulated values for $\zeta_c$ and $w_c$ for all alloys are shown in Table \ref{TableS1}, with very good agreement found.  Given the uncertainties in the simulations and the sinusoidal model, the agreement here broadly confirms the length scales controlling dislocation energetics in these HEA alloys.

\vspace*{-0.3cm}

\begin{table*}[h]
	\centering
	\caption{Values of $\lambda/4 = 1.4\zeta_c$ and $w_c$ as predicted by theory and as deduced from the correlation functions g(r) measured in the simulations.}
	\hspace*{-1.0cm}	\begin{tabular}{|c||c|c||c|c|}
		\hline
		Mo-Nb-Ta-V-W & $w_c$ theory (\AA) & $w_c$ simulations (\AA) & $\lambda/4$ theory (\AA) & $\lambda/4$ simulations (\AA) \\
		\hline
		\hline
		0.0-33.3-33.3-33.3-0.0 & 12.0 & 10.0 & 38.7 & 40.2 \\
		\hline
		21.7-20.6-15.6-21-21.1 & 12.0 & 8.4 & 48.9 & 49.4 \\
		\hline
		25.6-22.7-24.4-0.0-27.3 & 12.1 & 9.1 & 61.1 & 72.5 \\
		\hline
		24.9-25.8-26.6-22.7-0.0 & 12.1 & 11.0 & 45.6 & 52.4 \\
		\hline
		0.0-28.5-29.65-20.67-21.18 & 12.1 & 9.3 & 48.4 & 50.7 \\
		\hline
	\end{tabular}
	\label{TableS1}
\end{table*}

The predicted energy barriers for these alloys are all very high, $\Delta E_b \approx 1.9-2.5$ eV, leading to high predicted retained strengths at very high temperatures (see below).  These parameter-free results fully support the huge strengthening of, and thus the unexpected role of, edge dislocations in these BCC HEAs even at low temperatures.

\vspace*{-0.4cm}
\section{Theory {\it vs} experiments}

\vspace*{-0.2cm}
We now make predictions for Mo-Nb-Ta-V-W alloy compositions studied experimentally from 296K-1900K.  To do so, we must determine the solute/dislocation interaction energies $U_n(x_i, y_j)$ (negative when attractive).  We decompose the total interaction energy into an elastic misfit term plus a chemical interaction for sites near the highly distorted dislocation core, as
	\vspace*{-0.2cm}
	\begin{equation}
	U_n (x_i, y_j) = - p(x_i,y_j)\Delta V_n + U_n^{\mathrm{chem}}(x_i,y_j)
	\label{eq:Un-approx}
	\end{equation}
	
	\vspace*{-0.3cm}
Here, $p(x_i,y_j)$ is the pressure field generated by the dislocation structure in the average alloy matrix (e.g. Figure \ref{fig2}c).  

First-principles DFT is used to compute the solute misfit volumes in the true random MoNbTaW and MoNbTaVW alloys, as shown in Table \ref{Table1}.  The misfit volumes closely follow Vegard's law, $\Delta V_n = V_n - \bar{V}$ where $\bar{V} = \sum_{n=1}^N c_n V_n$ is the alloy atomic volume and {$V_n$} the elemental BCC atomic volumes.  The DFT-computed atomic volumes differ slightly from the experimental values, however.  So, we use Vegard's law based on the experimental atomic volumes of the alloy elements which yields values of $\bar{V}$ in good agreement with experiment.  As an aside, the misfit volumes computed using the Zhou et al. EAM potentials differ somewhat from the DFT values (see Table \ref{Table1}) and so are not used.  

The pressure field is computed using the dislocation core structure for the average alloy represented by the EAM potentials.  We have validated the EAM elastic constants $C_{11}$ and $C_{12}$  with respect to DFT-computed values for the MoNbTaW and MoNbTaVW alloys (Table \ref{Table1}).  The EAM value for $C_{44}$ is actually better than the DFT value, which is known to be underestimated in BCC metals~\cite{Koci2008}.  A rule-of-mixtures (ROM) estimate $\bar{C}_{ij} = \sum_{n=1}^N c_n C_{ij}$ using the EAM values of the elements yields good agreement with the atomistic values for the alloys and is used for other compositions.

The chemical energy for solutes near the core is then estimated using the EAM potentials.  We subtract the EAM elastic misfit energy (using the EAM misfit volumes) from the total EAM energy to obtain $U_n^{\mathrm{chem,EAM}} (x_i, y_j) = U_n^{\mathrm{EAM}}(x_i,y_j) - (- p(x_i,y_j)\Delta V_n^{\mathrm{EAM}})$.  Figure \ref{fig2}d shows $U_{\mathrm{Nb}}^{\mathrm{chem,EAM}}$ in NbTaV as an example of the magnitude and localization to atoms in the dislocation core.  

\vspace*{-0.2cm}
	\begin{table*}[h!!!]
	\small
	\centering
	\caption{Solute misfit volumes and elastic constants for the alloys studied, as computed using Density Functional Theory (data kindly provided by Dr. B. Yin~\cite{Yin2019a}, using methods detailed in Ref.~\cite{Yin2019b}), Vegard's law (misfits) or rule-of-mixtures ROM (elastic constants), the true random alloy described by EAM potentials, and the average-alloy EAM potential.}
	\hspace*{-2cm}	\begin{tabular}{c||c||c||c|c|c|c|c||c|c|c}
		\hline
		Mo-Nb-Ta-V-W & Method & $a_{\mathrm{bcc}}$  & $\Delta V_{\mathrm{Mo}}$ & $\Delta V_{\mathrm{Nb}}$ & $\Delta V_{\mathrm{Ta}}$ & $\Delta V_{\mathrm{V}}$ & $\Delta V_{\mathrm{W}}$ & $C_{11}$ & $C_{12}$ & $C_{44}$\\
		\hline
		\hline
		& DFT & 3.192 & -0.824 & 1.848 & 1.882 & -2.380 & -0.526 & 338 & 164 & 51 \\ \cline{2-11}
		\multirow{2}{*}{20-20-20-20-20} & Vegard/ROM & 3.192 & -0.628 & 1.713 & 1.877 & -2.484 & -0.478 & 346.8 & 157.7 & 90.5 \\ \cline{2-11}
		& EAM, Random & 3.201 & -0.956 & 1.246 & 1.571 & -1.547 & -0.333 & 306.3 & 156.8 & 79.1 \\ \cline{2-11}
		& EAM & 3.2 & -0.924 & 1.246 & 1.566 & -1.495 & -0.266 & 317.9 & 158.8 & 83 \\
		\hline
		\hline
		& DFT & 3.237 & -1.251 & 1.153 & 1.132 & --- & -1.034 & 374 & 163 & 64 \\ \cline{2-11}
		\multirow{2}{*}{25-25-25-0.0-25} & Vegard/ROM & 3.228 & -1.293 & 1.135 & 1.168 & --- & -1.010 & 375.5 & 167.3 & 101.6 \\ \cline{2-11}
		& EAM, Random & 3.223 & -1.263 & 1.014 & 1.162 & --- & -0.914 & 350.6 & 168.9 & 93.2 \\ \cline{2-11}
		& EAM & 3.221 & -1.218 & 1.019 & 1.181 & --- & -0.845 & 352.1 & 175.2 & 96.0 \\ 
		\hline
		\hline
		\multirow{2}{*}{21.7-20.6-15.6-21-21.1} & Vegard/ROM & 3.185 & -0.628 & 1.8 & 1.833 & -2.132 & -0.348 & 355.6 & 156.7 & 92.4 \\ \cline{2-11}
		\multirow{2}{*}{\scriptsize{(Nominal Mo-Nb-Ta-V-W alloy)}} & EAM, Random & 3.195 & -0.826 & 1.321 & 1.67 & -1.484 & -0.180 & 312.8 & 157.2 & 78.8 \\ \cline{2-11}
		& EAM & 3.194 & -0.803 & 1.316 & 1.653 & -1.434 & -0.127 & 317.9 & 158.8 & 83 \\
		\hline
		\hline
		\multirow{2}{*}{25.6-22.7-24.4-0.0-27.3} & Vegard/ROM & 3.224 & -1.229 & 1.199 & 1.232 & --- & -0.946 & 385.1 & 167.1 & 106 \\ \cline{2-11}
		\multirow{2}{*}{\scriptsize{(Nominal Mo-Nb-Ta-W alloy)}} & EAM, Random & 3.219 & -1.175 & 1.056 & 1.197 & --- & -0.846 &  357.5 & 170.6 & 96.2 \\ \cline{2-11}
		& EAM & 3.217 & -1.128 & 1.066 & 1.230 & --- & -0.774 & 358.8 & 174.4 & 97.7\\
		\hline
		\hline
		\multirow{2}{*}{24.9-25.8-26.6-22.7-0.0} & Vegard/ROM & 3.205 & -0.94 & 1.489 & 1.521 & -2.444 & --- & 300.8 & 146.6 & 72.8 \\ \cline{2-11}
		\multirow{2}{*}{\scriptsize{(Nominal Mo-Nb-Ta-V alloy)}} & EAM, Random & 3.211 & -1.194 & 1.205 & 1.615 & -1.962 & --- & 264.7 & 144.4 & 66.9 \\ \cline{2-11}
		& EAM & 3.21 & -1.156 & 1.215 & 1.627 & -1.886 & --- & 265.1 & 146.1 & 70.6 \\
		\hline
		\hline
		\multirow{2}{*}{0.0-28.5-29.65-20.67-21.18} & Vegard/ROM & 3.22 & --- & 1.258 & 1.29 & -2.675 & -1.171 & 310.3 & 152.5 & 78.2 \\ \cline{2-11}
		\multirow{2}{*}{\scriptsize{(Nominal Nb-Ta-V-W alloy)}} & EAM, Random & 3.231 & --- & 0.9457 & 1.218 & -2.046 & -0.969 & 268.5 & 150 & 71.5 \\ \cline{2-11}
		& EAM & 3.23 & --- & 0.95 & 1.205 & -1.95 & -0.938 & 267.7 & 149.3 & 75.6 \\
		\hline
	\end{tabular}
	\label{Table1}
\end{table*}

We study the experimentally-measured average (not nominal) compositions for each alloy.  All properties are shown in Table \ref{Table1}.  The uniaxial yield stress in an untextured BCC polycrystal controlled by edge dislocation glide is computed as $\sigma_y = M \tau_y$ (M=3.067).  The elastic moduli of Mo, Nb, Ta, V, and W are fairly insensitive to temperature up to 1900K (10-15\% decrease~\cite{Farraro1977,Farraro1979,Skoro2011}) and so we neglect temperature dependence of the alloy moduli.  We neglect possible strengthening effects due to grain-size and the actual dendritic/interdendritic as-cast microstructure (but see Appendix \ref{Supp3}).  
	
Figure \ref{fig3} shows very good agreement between predictions for the Mo-Nb-Ta-V-W alloys studied experimentally to date at T=296K and at the experimental strain rates.  There are no adjustable parameters in the predictions.  The theory rationalizes several features seen in the data.  First, the alloy with V is comparatively stronger because V has the largest misfit volume and so is the most potent strengthener.  Second, the alloys differing only by changing Mo to W have nearly the same strength because Mo and W misfit volumes are similar and alloy moduli changes are small.  The theory also predicts an activation volume $V_{act} \sim w_c \zeta_c b$ directly reflecting the underlying material length scales $w_c$ and $\zeta_c$.  For MoNbTaW at T=296K, we predict $V=36 b^3$, which happens to be in the range of other BCC alloys controlled by screw motion~\cite{Statham1972,Couzinie2015}.  Thus, even at low T in the region where edge and screw strengths may be comparable, the edge model provides good agreement with experiments.
		
As introduced earlier, Figures \ref{fig1} shows the predictions versus experiments for the MoNbTaW and MoNbTaVW alloys (at actual experimental compositions) versus temperature up to T=1900K.  The strength retention arises from the large zero-stress energy barriers ($\Delta E_b = 2.2$ eV, $2.5$ eV) created by the dislocation relaxation into a wavy low-energy structure in these random alloys.   The predictions are weakly dependent on the line tension constant $\alpha$, especially above 900K (Appendix \ref{Supp5}).  The experiments show a plateau in yield strength in the range 900K--1300K not predicted by the theory; one possible explanation is given in Appendix \ref{misc}.  These results show that the edge strengthening agrees fairly well with the measured high temperature behavior.  The predictions are lower than experiments.  However, the factors of grain-size Hall-Petch strengthening, solute-solute interactions, cast microstructure, and impurities can all increase the overall strength.  Temperature-independent Hall-Petch strengthening can be estimated as $\approx 50$MPa from data on the elemental metals.  Solute-solute interactions exists in MoNbTaW and have been estimated in MoNbTaW using DFT~\cite{Kormann2017}; these will also increase the strength and barrier.  Impurities such as O and N also increase strength and barriers.  Thus theory predictions purely based on the critical resolved shear stress can be lower than experimental measurements of the yield strength.

\section{Analytic model based on elasticity}

Application of the full model requires detailed information about the solute-dislocation interaction energies $U_n(x_i, y_j)$.  This information is not accessible experimentally and is very difficult to obtain using first-principles methods.  The Zhou et al. EAM potentials have proved to be rather good for the systems studied to date, but may not be quantitatively accurate for all systems.  It is thus beneficial to seek a simplified theory that can be easily applied while retaining good accuracy.   Such a theory is achieved, as previously done for FCC alloys, by using the elasticity approximation $U_n(x_i, y_j) = -p(x_i, y_j)  \Delta V_n$.  The dislocation pressure field $p(x_i, y_j)$ can be written as \linebreak $p(x_i, y_j) = -\frac{\bar{\mu}}{3 \pi} \frac{(1+\bar{\nu})}{(1-\bar{\nu})} f(x_i, y_j)$ where $f(x_i, y_j)$ is a dimensionless anisotropic pressure field generated by the distribution of normalized Burgers vector along the glide plane (see Fig. \ref{fig2}c) with the isotropic alloy elastic constants $\bar{\mu}$ and $\bar{\nu}$ introduced for scaling .  Inserting the elasticity approximation into Eq.~(\ref{eq:energyscale}), the key energy in the theory becomes
\vspace*{-0.2cm}
\begin{equation}
\Delta \tilde{E}_p(w)  = \frac{\bar{\mu}}{3\pi}\frac{(1+\bar{\nu})}{(1-\bar{\nu})}\left[\sum_{i,j}\Delta f_{ij}^2(w)\right]^{\frac{1}{2}} \times \left[\sum_n c_n \Delta V_n^2 \right]^{\frac{1}{2}},
\label{eq:DE_p_tilde_pdVm} 
\end{equation}
where $\Delta f_{ij}(w) = f(x_i-w,y_j)-f(x_i,y_j)$.
The quantity $\sum_n c_n \Delta {V}_n^2$ emerges as the crucial misfit volume quantity.  The minimization to obtain $w_c$ is then determined only by the dislocation core structure through the quantity $\Delta f_{ij}(w)$, independent of the solute properties.  This is a very revealing result, due to its generality and separation of the problem into misfit volumes and, independently, the dislocation core structure of the average matrix.  However, it is not necessarily quantitatively accurate.

\begin{figure*}[h!!]
	\centering
	\hspace*{-0.7cm}
	\includegraphics[width=17 cm]{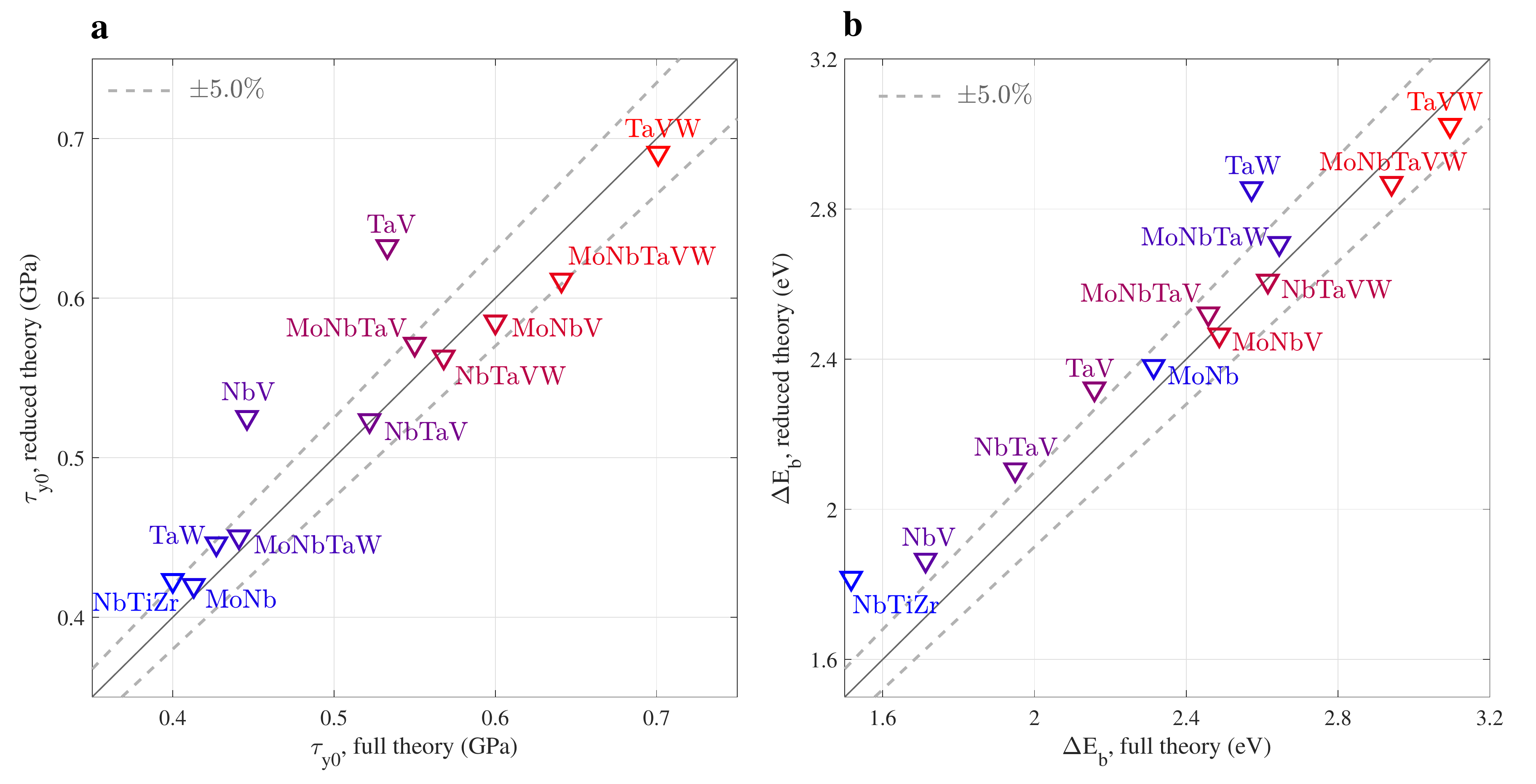}
	\vspace*{-0.5cm}
	\caption{{\bf Reduced {\it vs} full theory.} \textbf{a} predictions of $\tau_{y0}$ for a number of alloys and \textbf{b} predictions of $\Delta E_b$ for the same alloys.}
	\label{fig6}
\end{figure*}

After minimization with respect to $w$, the elasticity theory enables the key theory quantities to be expressed as
\vspace*{-0.4cm}
\begin{equation}
\label{eq:tauy0_pdVm}
\tau_{y0}  = 0.051 \alpha^{-\frac{1}{3}} \bar{\mu}\left(\frac{1+\bar{\nu}}{1-\bar{\nu}}\right)^{\frac{4}{3}} f^{\tau} \times \left[\frac{\sum_n c_n \Delta V_n^2}{b^6} \right]^{\frac{2}{3}}\ ,
\end{equation}
\begin{equation}
\Delta E_b  =  0.274 \alpha^{\frac{1}{3}}\bar{\mu} b^3 \left(\frac{1+\bar{\nu}}{1-\bar{\nu}}\right)^{\frac{2}{3}}f^{\Delta E} \times \left[\frac{\sum_n c_n \Delta V_n^2}{b^6}\right]^{\frac{1}{3}}\ .
\label{eq:DEb_pdVm}
\end{equation}
where $f^{\tau}$ and $f^{\Delta E}$ are dimensionless constants emerging from the minimization and related to the normalized pressure field of the dislocation.  We have used the above form coming from elasticity to obtain {\it effective} parameters  $f^{\tau}$ and $f^{\Delta E}$ by fitting Eqs. (\ref{eq:tauy0_pdVm},\ref{eq:DEb_pdVm}) to the full results obtained using Eqs. (\ref{eq:DE_b},\ref{eq:tau_y0})  for many different equicomposition binary, medium- and high-entropy alloys with Eqs. (\ref{eq:DE_b},\ref{eq:tau_y0}).  The fitting to achieve single {\it effective} values of $f^{\tau}$ and $f^{\Delta E}$ across all alloys corresponds to assuming a constant core structure for all alloys and an embedding of the chemical energies into the form of the elasticity theory.  The strength and energy barrier for any alloy are now given by
\vspace*{-0.4cm}
\begin{equation*}
\tau_{y0} = 0.040 \alpha^{-\frac{1}{3}} \bar{\mu} \left( \frac{1+\bar{\nu}}{1-\bar{\nu}}\right)^{\frac{4}{3}} \left[ \frac{\sum_n c_n \Delta V_n^2}{\overline{b}^6} \right]^{\frac{2}{3}}
\end{equation*}
\begin{equation*}
\Delta E_b = 2.00 \alpha^{\frac{1}{3}} \bar{\mu} \overline{b}^3 \left( \frac{1+\bar{\nu}}{1-\bar{\nu}}  \right)^{\frac{2}{3}} \left[ \frac{\sum_n c_n \Delta V_n^2}{\overline{b}^6} \right]^{\frac{1}{3}}
\label{eq:elasticity}
\end{equation*}
where $\bar{\mu},\bar{\nu}$ are the isotropic alloy elastic constants and $\overline{b}$ is the alloy Burgers vector, calculated using Vegard's law to determine the alloy volume.  In the above, the alloy elastic moduli are computed as before ($\bar{\mu} =\sqrt{\frac{1}{2}\bar{C}_{44}\bar{C}_{11}-\bar{C}_{12})}$; $\bar{B} = (\bar{C}_{11} + 2\bar{C}_{12})$; $\bar{\nu} = \frac{3\bar{B}-2\mu}{2(3\bar{B}+\mu)}$).  The analytic formulae above have been fitted to reproduce most high-strength/high-temperature data within 5\% accuracy, and differ from the whole dataset of the full calculations (Eqs.~(\ref{eq:DE_b},\ref{eq:tau_y0},\ref{eq:Un-approx})) with standard deviation of $\sim$10\% for both strength and energy barrier.  The comparison between full theory and reduced theory predictions are shown in Figure \ref{fig6}.  This simplified analytic theory thus depends only on elastic moduli and misfit volumes.  Combined with Eqs.~(\ref{eq:tau_y_T_epsdot2},\ref{eq:finiteT_tay_y2}), they can be easily applied to provide guidance on the composition-dependence of yield strength versus temperature.

\vspace*{-0.4cm}

\section{Alloy discovery}
	
The simplified elasticity-based edge theory can now be used to search for other high-strength and high-strength/weight compositions in the Mo-Nb-Ta-V-W family.

\begin{figure*}[h!!]
	\centering
	\vspace*{-2cm}
	\hspace*{-1.5cm}
	\includegraphics[width=19 cm]{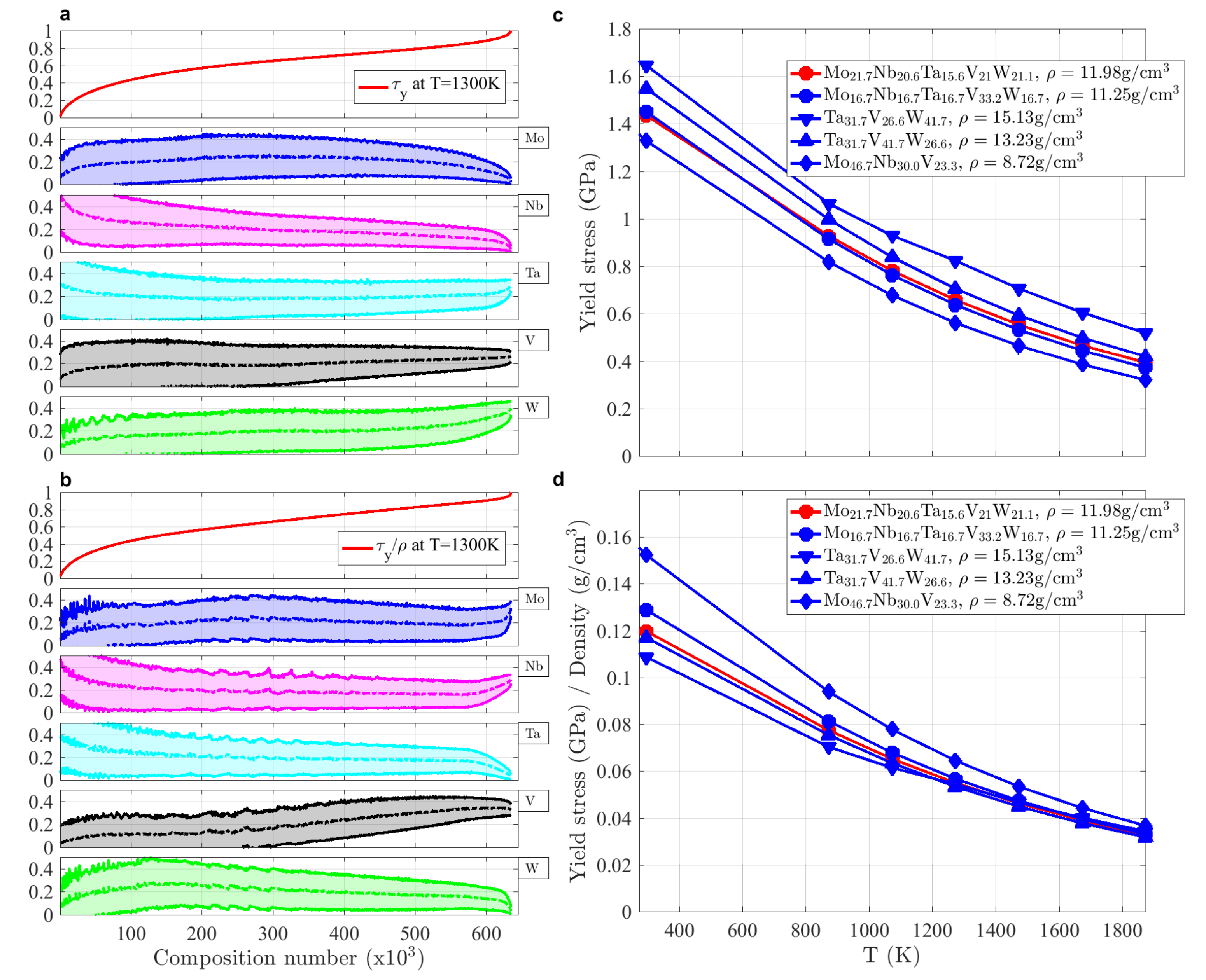}
	\vspace*{-0.5cm}
	\caption{{\bf Theory predictions for compositions emerging from the optimization process.} \textbf{a} Theory predictions for strength at T=1300K {\it vs} composition (average $\pm$ standard deviation for 1000 compositions per bin), normalized by the highest strength found in the optimization model. \textbf{b} Theory predictions for strength/density {\it vs} composition (average $\pm$ standard deviation for 1000 composition per bin), normalized by the highest strength/density found in the optimization model. \textbf{c} Theory predictions of strength {\it vs} versus temperature over 296K--1900K for several near-optimal compositions, with the experimental and predicted results on the existing 5-element MoNbTaVW alloy~\cite{Senkov2011} also shown (in red). \textbf{d} Theory predictions of strength/density {\it vs} versus temperature over 296K--1900K for near-optimal compositions, with the experimental and predicted results on the existing 5-element MoNbTaVW alloy~\cite{Senkov2011} also shown (in red).}
	\label{fig4}
\end{figure*}

With inputs obtained from the accurate Vegard's law for misfit volumes and rule-of-mixtures for elastic moduli, which require only simple algebra, a search over the entire composition space can be easily executed.  We have considered the entire composition space using over $> 6 \times 10^5$ compositions differing by at least 1.6at.$\%$ (each elemental concentration varied from 0 to 1 in increments of 1/60).  The predicted strengths and strength/weight ratios at T=1300K are shown in Figures \ref{fig4}a,b normalized by the top-performing alloy values in each case.  Thousands of compositions are within $\sim 10\%$ of the strongest alloy.  Many alloys near the maximum strength/weight are also near the maximum strength.   The trends are clear: high strength is achieved with combinations of small and large elements (e.g. V and Ta) and/or high-stiffness elements (e.g. W).  High strength/weight is achieved by using the lighter atoms (Mo, Nb, V) but in compositions that give the largest misfit parameters and/or elastic constants.

From the analytic model results, we have selected the estimated highest-strength alloys at T=1300K in the Ta-V-W, Mo-Nb-Ta-V-W, and Mo-Nb-V families and performed calculations using the full model.  The results for these selected alloys are shown in Figures \ref{fig4}c,d.   We identify new alloys that are predicted to have higher T=1300K strength (Ta$_{31.7}$V$_{26.6}$W$_{41.7}$, Ta$_{31.7}$V$_{41.7}$W$_{26.6}$) or higher strength/weight ratio (Mo$_{46.7}$Nb$_{30.0}$V$_{23.3}$) than the existing 5-component alloy.  The 5-component alloy is the strongest alloy measured to date at T=1000C \cite{Senkov2018}, and so the predicted new alloys hold promise for being among the strongest alloys possible in this system.   Such guidance on attractive compositions, achieved with minimal computational effort, is powerful demonstration of the power of the theoretical framework for property prediction in such complex alloys.   With these predictions, we await experimental fabrication and testing of some of these alloys.

\section{Discussion and summary}

Unlike superalloys (Figure \ref{fig1}), the MoNbTaW and MoNbTaVW BCC HEAs do not show a precipitous drop in strength at high T (roughly half the melting temperature $T_m$).  The present edge strengthening mechanism rationalizes this result.  The strength is {\it intrinsic} to the nature of the atomic-scale complexity of the HEA alloy and does not rely on mechanisms that can be easily defeated by high-temperature diffusional/dislocation-climb processes.  Thus, the dislocations must move through the random alloy itself - there are no easy paths of dislocation motion that can circumvent the large barriers, which appear throughout the alloy on the scales of $\zeta_c$ and $w_c$, responsible for the strengthening.  In contrast, the high-temperature strength of the screw dislocation is dominated by pinning due to jogs.  Jog strengthening is defeated at high temperatures by thermal vacancies that eliminate the high energy barrier for self-interstitial creation that controls the jog strength.  Thus, the differences in high temperature performance among different BCC HEAs - the pressing issue identified by \cite{Senkov2018} - may be directly attributable to screw versus edge control of strengthening.  The edge theory alone also provides a lower-bound estimate of the high-T strengthening.  That is, if the edge theory predicts a high strength, then a high strength can be expected.  If the edge theory predicts a low strength, then the alloy may be screw controlled, and still have good high-T properties up to roughly $0.5 T_m$. 

The prediction of a wide range of compositions that match or exceed the strengths of existing alloys (Figure \ref{fig4}) opens avenues for optimization across a much broader range of properties such as oxidation resistance, diffusional creep, and ductility, while maintaining high-temperature strength.  With suitable models for those properties, optimization can be performed to discover alloys having the desired mix of properties.  Mechanical property optimization can also be combined with thermodynamic models to avoid compositions where undesirable intermetallic phase formation is predicted~\cite{delGrosso2012,Huhn2013,Kormann2017,Wang2018}, thus allowing for simultaneous computationally-guided design of performance and processing of new alloys.  Together with an associated model for strengthening of screw dislocations \cite{Maresca2018b}, the models present a framework for analyzing BCC HEAs across all families and all temperatures.	

In summary, a mechanistic parameter-free theory quantitatively captures the exceptional high flow stresses from T=0K-1900K of Mo-Nb-Ta-V-W High Entropy Alloys.  The trapping of dislocations in statistically-favorable random solute environments creates intrinsic large energy barriers for the edge dislocation motion.  The nature of this trapping of the dislocations makes these alloys robust against typical high temperature softening mechanisms.  A reduced analytic version of the theory enables rapid screening across the entire composition space, leading to the identification of new promising alloys.  This opens avenues for computationally-guided multi-property optimization and discovery of new high-performance materials.

\normalstyle
	
	\subsection*{Data availability}
	
	For access to more detailed data than are given in the article or the Appendix please contact the authors.
	
	
	

\begin{thebibliography}{100}

	\makeatletter
	\renewcommand\@biblabel[1]{#1.}
	\makeatother

		
		

		
		
		\bibitem{Senkov2010} Senkov, O.~N., Wilks, G.~B., Miracle, D.~B., Chuang, C.~P., Liaw, P.~K. Refractory high-entropy alloys. {\it Intermetallics} {\bf 18}, 1758--1765 (2010).
		
		\bibitem{Senkov2011}
		Senkov, O.~N., Wilks, G.~B., Scott, J.~M., \& Miracle, D.~B. Mechanical properties of Nb25Mo25Ta25W25 and V20Nb20Mo20Ta20W20 refractory high entropy alloys. {\it Intermetallics} {\bf 19}, 698--706 (2011).
		
		\bibitem{Yao2016a}
		Yao, H.~W. {\it et al.} NbTaV-(Ti,W) refractory high-entropy alloys: Experiments and modeling. {\it Mater. Sci. Eng. A} {\bf 674}, 203--211 (2016).
		
		\bibitem{Yao2016b}
		Yao, H.  {\it et al.} MoNbTaV medium-entropy alloy. {\it Entropy} {\bf 18}, 1--15 (2016).
		
		\bibitem{Gludovatz2014}
		Gludovatz, B.~ {\it et al.} Ritchie, A fracture-resistant high-entropy alloy for cryogenic applications. {\it Science} {\bf 345}, 1153--1158 (2014).
		
		
		\bibitem{Wu2014a}
		Wu, Z., Bei, H., Pharr, G., George, E. Temperature dependence of the mechanical properties of equiatomic solid solution alloys with face-centered cubic crystal structures. {\it Acta Mater.} {\bf 81}, 428--441 (2014).
		
		
		\bibitem{Li2016}
		Li, Z., Pradeep, K.~G., Deng, Y., Raabe, D., Tasan, C.~C. Metastable high-entropy dual-phase alloys overcome the strength-ductility trade-off. {\it Nature} {\bf 534}, 227--230 (2016).
		
		\bibitem{Miracle2017}
		Miracle, D.~B., Senkov, O.~N. A critical review of high entropy alloys and related concepts. {\it Acta Mater.} {\bf 122}, 448--511 (2017).
		
		\bibitem{Senkov2018} Senkov, O.~N., Miracle, D.~B., Chaput, K.~J., Couzini\'{e}, J.-Ph. Development and exploration of refractory high entropy alloys -- A review. {\it J. Mater. Res.} {\bf 33}, 3092--3128 (2018).
		
		\bibitem{Rodney2014}
Rodney, D., Bonneville, J. Dislocations. {\it In Physical Metallurgy}, Elsevier Oxford, 2014.

\bibitem{Cordero2016}
Cordero, Z.~C., Knight, B.~E., Schuh, C.~A. Six decades of the Hall-Petch effect - a survey of grain-size strengthening studies on pure metals. {\it Int. Mater. Rev.} {\bf 61}, 495--512 (2016).

\bibitem{Suzuki1979}
Suzuki, H. Solid Solution Hardening in Body-Centred Cubic Alloys. {\it In Dislocations in Solids}, North-Holland, 1979.

\bibitem{Trinkle2005}
Trinkle, D.~R., Woodward, C. The chemistry of deformation: How solutes soften pure metals. {\it Science} {\bf 310}, 1665--1667 (2005).

\bibitem{Maresca2018b} Maresca, F., Curtin, W.~A. Theory of screw dislocation strengthening in random BCC alloys from dilute to ``High-Entropy'' alloys. {\it Submitted} (2019).

\bibitem{Statham1972}
Statham, C.~D., Koss, D.~A., Christian, J.~W. The thermally activated deformation of Niobium-Molybdenum and Niobium-Rhenium alloy single crystals. {\it Phil. Mag.} {\bf 26}, 1089--1103 (1972).

\bibitem{Caillard2013} Caillard, D. A TEM in situ study of alloying effects in iron. II--Solid solution hardening caused by high concentrations of Si and Cr. {\it Acta Mater} {\bf 61}, 2808--2827 (2013).

\bibitem{Dirras2015} Dirras, G. {\it et al.} Microstructural investigation of plastically deformed Ti$_{\mathrm{20}}$Zr$_{\mathrm{20}}$Hf$_{\mathrm{20}}$Nb$_{\mathrm{20}}$Ta$_{\mathrm{20}}$ high entropy alloy by X-ray diffraction and transmission electron microscopy. {\it Mater. Char.} {\bf 108}, 1--7 (2015).

\bibitem{Couzinie2015}
Couzini\'{e}, J.-Ph.  {\it et al.}, On the room temperature deformation mechanisms of a TiZrHfNbTa refractory high-entropy alloy. {\it Mater. Sci. Eng. A} {\bf 645}, 255--263 (2015).

\bibitem{Couzinie2018} Couzini\'{e}, J.-Ph., Dirras, G., Mompiou, F., Caillard, D., Guillot, I. Body-centered cubic high-entropy alloys -- Understanding of the mechanical properties and associated underlying deformation mechanisms. {\it MRS Fall Meeting}, Boston (2018).

\bibitem{Mompiou2018} Mompiou, F., Tingaud, D., Chang, Y., Gault, B., Dirras, G. Conventional vs harmonic-structured $\beta$-Ti-25Nb-25Zr alloys: A comparative study of deformation mechanisms. {\it Acta Mater.} {\bf 161}, 420--430 (2018).




		\bibitem{Yao2017}
		Yao, H.~W. {\it et al.}, Mechanical properties of refractory high-entropy alloys: Experiments and modeling. {\it J. Alloys Compd.} {\bf 696}, 1139--1150 (2017).
		

		
		
		
		
		
		
		\bibitem{Chen2018}
		Chen, H.  {\it et al.} Contribution of lattice distortion to solid solution strengthening in a series of refractory high entropy alloys. {\it Metall. Mater. Trans. A} {\bf 49}, 772--781 (2018).
		
\bibitem{Argon2007}
Argon, A. Strengthening Mechanisms in Crystal Plasticity, Oxford Univ. Press,
2007.

		\bibitem{Dirras2017} Dirras, G., Tingaud, D., Ueda, D., Hocini, A., Ameyama, K. Dynamic Hall-Petch {\it versus} grain-size gradient effects on the mechanical behavior under simple shear loading of $\beta$-titanium Ti-25Nb-25Zr alloys. {\it Mater. Lett.} {\bf 206}, 214--216 (2017).


		
		\bibitem{Varvenne2016}
		Varvenne, C., Luque, A., Curtin, W.~A. Theory of strengthening in fcc high entropy alloys. {\it Acta Mater.} {\bf 118}, 164--176 (2016).
		
		\bibitem{Leyson2012} Leyson, G.~P.~M., Hector Jr., L.~G., Curtin, W.~A. Solute strengthening from first principles and application to aluminum alloys. {\it Acta Mater.} {\bf 60}, 3873--3884.
		
		\bibitem{Kocks1975}
		Kocks, U., Argon, A.~S., Ashby, M.~F. Models for macroscopic slip. {\it Prog. Mater. Sci.} {\bf 19}, 1--281 (1975).
		
		\bibitem{Leyson2009} Leyson, G.~P.~M., Curtin, W.~A. Solute strengthening at high temperatures. {\it Model. Simul. Mater. Sci. Eng.} {\bf 24}, 065005 (2016).
		
		\bibitem{Plimpton1995} Plimpton, S. Fast parallel algorithms for short-range molecular dynamics. {\it J.  Comput. Phys.} {\bf 117}, 1--19 (1995).
		
		\bibitem{Zhou2004}
		Zhou, X.~W., Johnson, R.~A., Wadley, H.~N.~G. Misfit-energy-increasing dislocations in vapor-deposited CoFe/NiFe multilayers. {\it Phys. Rev. B} {\bf 69}, 1--10 (2004).
		
		\bibitem{Lin2013} Lin, D.-Y., Wang, S.~S., Peng, D.~L., Li, M., Hui, X.~D. An {\it n}-body potential for a Zr-Nb system based on the embedded-atom method. {\it J. Phys.: Condens. Matter} {\bf 25} 105404 (2013).
		
		\bibitem{Rao2017}
		Rao, S.I. {\it Unpublished research}.
		
		\bibitem{Varvenne2016b}
		Varvenne, C., Luque, A., N\"{o}hring, W.~G., Curtin, W.~A. Average-atom interatomic potential for random alloys. {\it Phys. Rev. B} {\bf 93}, 104201 (2016).

		\bibitem{Bacon2009} Bacon, D.~J., Osetsky, Y.~N., Rodney, D. Dislocation-obstacle interactions at the atomic level. {\it Dislocations in Solids} (Eds. J.~P. Hirth, L. Kubin) {\bf 15}, 1--90 (2009).

		\bibitem{Bitzek2006} Bitzek, E., Koskinen, P., G\"{a}hler, F., Moseler, M., Gumbsch, P. Structural relaxation made simple. {\it Phys. Rev. Letters} {\bf 97}:170201 (2006).
		
		\bibitem{Koci2008} Ko\v{c}i, L., Ma, Y., Oganov, A.~R., Souvatzis, P., Ahuja, R. Elasticity of the superconducting metals V, Nb, Ta, Mo, and W at high pressure. {\it Phys. Rev. B} {\bf 77}, 214101 (2008).
		
		\bibitem{Stukowski2010} Stukowski, A. Visualization and analysis of atomistic simulation data with OVITO, the Open Visualization Tool. {\it Model. Simul. Mater. Sci. Eng.} {\bf 18}, 015012 (2009).
		
		\bibitem{Stukowski2010b} Stukowski, A., Albe, K. Extracting dislocations and non-dislocation crystal defects from atomistic simulation data. {\it Model. Simul. Mater. Sci. Eng.} {\bf 18}, 085001 (2010).
		
		\bibitem{Yin2019a} Yin, B., {\it private communication}, (2019).

		\bibitem{Yin2019b} Yin, B., Curtin, W.~A. First-principles-based prediction of yield strength in the RhIrPdPtNiCu high-entropy alloy. {\it npj Comput. Mater.} {\bf 5}, 14 (2019).
		
		\bibitem{Kormann2017}
		K\"{o}rmann, F., Ruban, A.~V., Sluiter, M.~H.~F. Long-ranged interactions in bcc NbMoTaW high-entropy alloys. {\it Mater. Res. Lett.} {\bf 5}, 35--40 (2017).
					
		\bibitem{Farraro1977}
		Farraro, R.~J., McLellan, R.~B. Temperature dependence of the Young's modulus and shear modulus of pure nickel, platinum, and molybdenum. {\it Metall. Trans. A} {\bf 8}, 1563--1565  (1977).
		
		\bibitem{Farraro1979}
		Farraro, R.~J., McLellan, R.~B. High temperature elastic properties of polycrystalline niobium, tantalum, and vanadium. {\it Metall. Trans. A} {\bf 10}, 1699--1702 (1979).
		
		\bibitem{Skoro2011}
		\v{S}koro, G.~P. {\it et al.}, Dynamic Young's moduli of tungsten and tantalum at high temperature and stress. {\it J. Nucl. Mater.} {\bf 409}, 40--46 (2011).
		
		\bibitem{delGrosso2012}
		del Grosso, M.~F., Bozzolo, G., Mosca, H.~O. Determination of the transition to the high entropy regime for alloys of refractory elements. {\it J. Alloys Compd.} {\bf 534}, 25--31 (2012).
		
		\bibitem{Huhn2013}
		Huhn, W.~P., Widom, M. Prediction of A2 to B2 phase transition in the high-entropy alloy Mo-Nb-Ta-W. {\it JOM} {\bf 65}, 1772--1779 (2013).
		
		\bibitem{Wang2018}
		Wang, Y. {\it et al.} Computation of entropies and phase equilibria in refractory V-Nb-Mo-Ta-W high-entropy alloys. {\it Acta Mater.} {\bf 143}, 88--101 (2018).
		
		\bibitem{Varvenne2017} Varvenne, C., Leyson, G.~P.~M., Ghazisaeidi, M., Curtin, W.~A. Solute strengthening in random alloys. {\it Acta Mater.} {\bf 124}, 660--683 (2017).		

		\bibitem{Szajewski2015} Szajewski, B., Pavia, F., Curtin, W.~A. Robust atomistic calculation of dislocation line tension. {\it Model. Simul. Mater. Sci. Eng.} {\bf 23}, 085008 (2015).	
		
		\bibitem{Curtin2006} Curtin, W.~A., Olmsted, D.~L., Hector~Jr., L.~G. A predictive mechanism for dynamic strain ageing in aluminium-magnesium alloys. {\it Nature Mater.} {\bf 5}, 875--880 (2006).	
		
		
				
		
		
		
		
		
		
		
		
		
		
	\end{thebibliography}
	



\section*{Acknowledgments}

Support for this work was provided through a European Research Council Advanced Grant, ``Predictive Computational Metallurgy", ERC Grant agreement No. 339081 - PreCoMet.  Computational resources were supported by EFPL funding to the LAMMM lab and executed through the EPFL SCITAS HPC facility.  The authors thank Dr. Binglun Yin for providing the DFT computations of elemental and alloy properties.

\section*{Contributions}

F.M. and W.A.C. designed the research, analyzed the data, developed the model, discussed the results, and wrote the paper. F.M. performed the Molecular Dynamics simulations.

\section*{Competing Interests}

The authors declare no competing financial interests.

\section*{Material and Correspondence}

Correspondence and materials requests should be addressed to F.M. (francesco.maresca@epfl.ch)

\appendix

\section{Dependence of theory predictions on dislocation line tension} \label{Supp5}

The T=0K yield strength and energy barrier depend on the line tension $\Gamma$, scaling as  $\tau_{y0} \approx \Gamma^{-1/3}$ and $\Delta E_b \approx \Gamma^{1/3}$, respectively.  Changes in $\Gamma$ thus change the strength and barrier in opposite directions, leading to some cancellation of effects at moderate and high temperatures.  Thus, predictions are not strongly sensitive to $\Gamma$ except at very low temperature.

The line tension is generally expressed as $\Gamma = \alpha \bar{\mu}b^2$, which captures the proper scalings with (effective isotropic) alloy shear modulus $\bar{\mu}$ and $b$.  Variations in $\Gamma$ are thus manifest through the non-dimensional parameter $\alpha$.  Work in FCC alloys has previously used $\alpha = 1/8$ (\cite{Varvenne2016,Varvenne2017}) based on atomistic simulations of dislocation bow-out~\cite{Szajewski2015}. Varvenne et al.$^{27}$ have shown that the low-temperature strength may be better-predicted using the value $\alpha = 1/16$.  For long dislocation lines, the line tension is dominated by elasticity and so is independent of any underlying crystal structure.  Thus, aside from the contributions due to dislocation core energy that are not negligible at very small dislocation lengths, values for $\alpha$ for BCC metals are expected to be in the same range as those for FCC metals.  

\begin{figure*}[h!!]
	\centering
	\hspace*{-2cm}
	\includegraphics[width=20 cm]{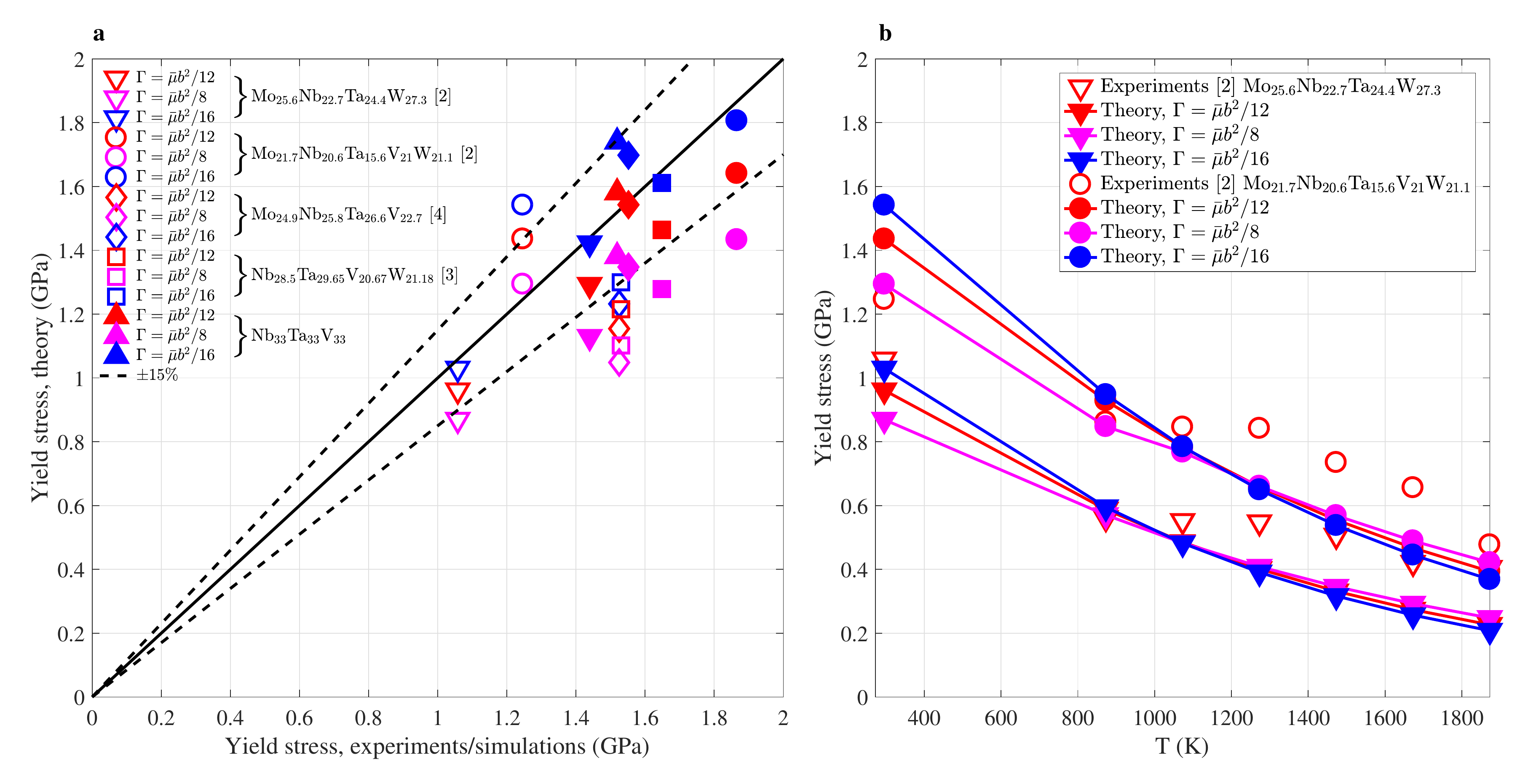}
	\vspace*{-1.2cm}
	\caption{{\bf Effect of line tension on theory predictions.} \textbf{a} Theory predictions {\it vs} T=0K simulations and RT experiments, for line tension parameters $\alpha = 1/16, 1/12, 1/8$ as indicated.  \textbf{b} Theory predictions {\it vs} temperature from 296K--1900K, with experimental strengths also shown for reference.  All other details are identical to those used in the main text.}
	\label{figSI5_1}
\end{figure*}

With the above background, Figure \ref{figSI5_1} shows the predictions for strengths of the BCC HEAs for the values $\alpha = 1/16, 1/12, 1/8$ that span the expected range.  Results in the main text use $\alpha = 1/12$.  The results in Figure \ref{figSI5_1}a show that the room temperature (RT) predictions are rather sensitive to the specific line tension. However, as anticipated, the results in Figure \ref{figSI5_1}b show that predictions at moderate to high temperatures (296--1900K) are quite insensitive to the specific choice of the line tension $\Gamma$, and within the uncertainties associated with other details (grain size effects, as-cast microstructure effects, solute/dislocation interaction energies, model uncertainty).
\vspace*{-0.5cm}

\section{Effects of microstructure and compositional homogeneity} \label{Supp3}

No grain-size Hall-Petch effect has been accounted introduced here.  However, while the average grain size of the as-cast materials is fairly large, $80-200\ \mu m$, there is likely some modest H-P effect.  An empirical rule-of-mixtures relation can be used along with the measured Hall-Petch effects in the elemental metals$^{15}$, resulting in a temperature-independent increase of the yield strength by $\sim 50$ MPa.

\begin{figure*}[h!!]
	\centering
	\hspace*{-2cm}
	\includegraphics[width=20 cm]{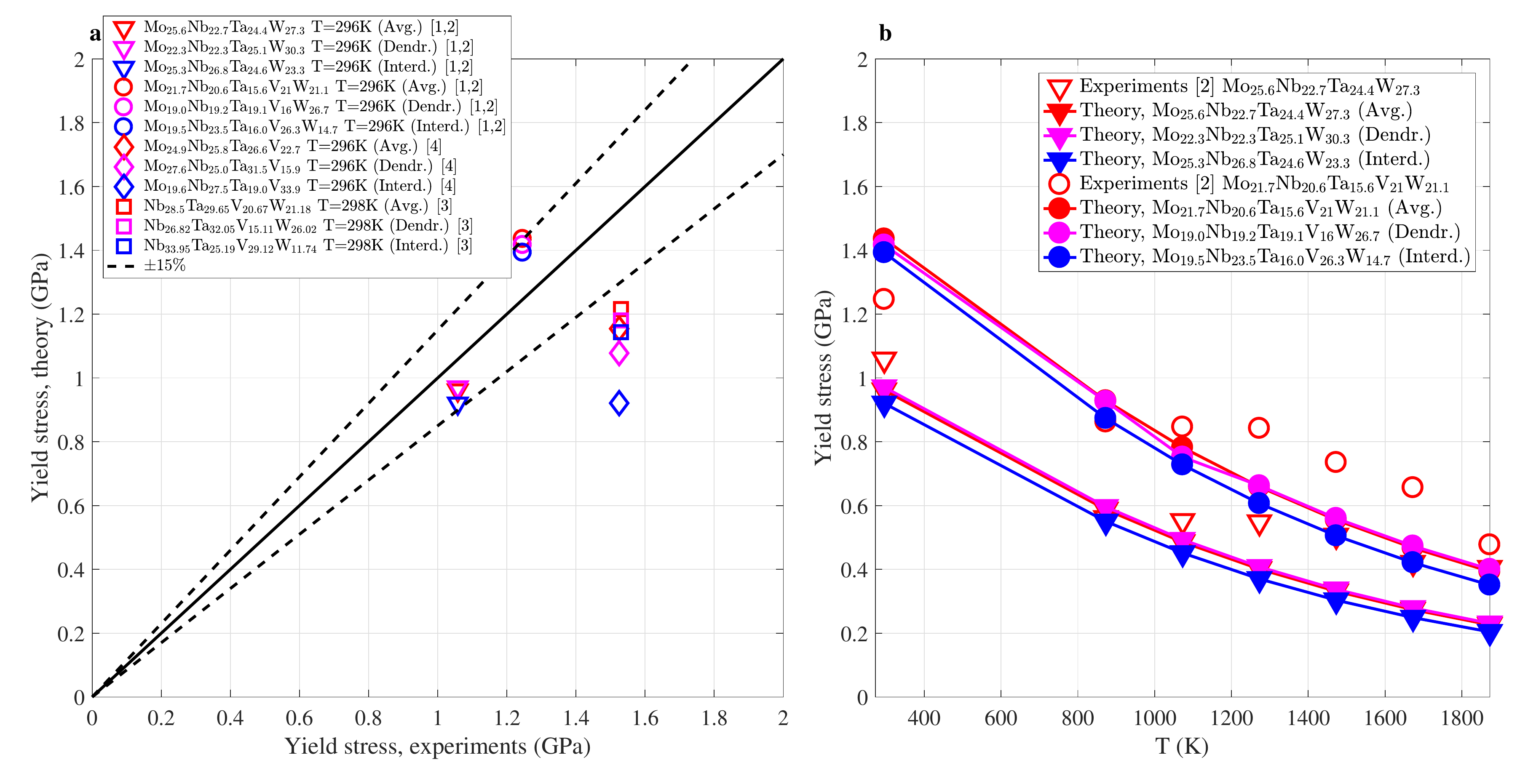}
	\caption{{\bf Effect of as-cast local composition fluctuations on theory predictions.} \textbf{a} Theory predictions for the average, dendritic and interdendritic compositions {\it vs} RT experiments. \textbf{b} Theory predictions for the average, dendritic and interdendritic compositions {\it vs} experiments from 296K--1900K.}
	\label{SI12}
\end{figure*}

Results shown in the main text also relate to the average compositions measured in the as-cast specimens.  The as-cast speciments have a dendritic structure at the scale of $\approx 20 \mu m$, with different compositions in the dendritic and interdendritic regions.  If the lattice constants and yield strengths of these two regions are very similar, then the overall composite material may behave as an effectively single-phase material. To investigate this issue, we have applied the theory to predict the strengths of the dendritic and interdendritic materials using their reported average compositions, as shown in  Figure \ref{SI12}.

For the alloys considered here, except MoNbTaV, the values closely straddle the value of the overall average composition.  For the MoNbTaV alloy, the dendritic region is 158 MPa stronger than the interdendritic composition, and 78 MPa weaker than the average composition.  The strength of the experimental MoNbTaV alloy may thus be between the dendritic and interdendritic values (closer to dendritic since the phase fraction of dendrites is higher than interdendritic regions).  Limited data is available on annealed materials.  Annealing of a HfMoNbTiZr BCC alloy$^{24}$ led to a decrease in strength of 144 MPa.  Similar decreases may apply to other materials.   On the other hand, the small size of the dendritic arms ($20\ \mu m$) might provide further Hall-Petch-type strengthening. These aspects may explain why the theory underpredicts the measured strength in MoNbTaV, although the predictions remain quite reasonable.

\section{Possible origin of the high-T strength plateau} \label{misc}

The experiments show a plateau in yield strength in the range 800K--1100K. The origin of this plateau in BCC alloys is not well-established.  We postulate that dynamic strain aging (DSA) via ``cross-core diffusion"~\cite{Curtin2006} may generate additional strengthening in these alloys at intermediate temperatures.  Cross-core diffusion occurs when solutes diffuse locally \textit{only} across the core from higher-energy sites to lower-energy sites (see Figure \ref{fig2}b).  This leads to a time- and temperature-dependent strengthening that saturates once all cross-core motion has occurred, after which the normal decrease in strength with increasing temperature resumes.  This is consistent with the experimental results in Figure \ref{fig2}b.  The origins of the plateau require deeper study but do not detract from the broad success of the theory.

\end{document}